\documentclass[%
 reprint,
superscriptaddress,
 amsmath,amssymb,
 aps,pre,hyphens,floatfix
]{revtex4-2}

\usepackage[]{amsmath}
\usepackage{amssymb}
\usepackage{color}
\usepackage{enumerate}
\usepackage[usenames,dvipsnames,svgnames,table]{xcolor}
\usepackage{IEEEtrantools}
\usepackage{graphicx}
\usepackage{array}
\usepackage{multirow}
\usepackage{wrapfig}
\usepackage{fullpage}
\usepackage{overpic}
\usepackage{verbatim}
\usepackage{float}
\usepackage{neuralnetwork}
\usepackage[caption=false]{subfig}
\usepackage{multirow}
\usepackage{diagbox}
\usepackage[normalem]{ulem}  

\begin{document}

\preprint{APS/123-QED}

\title{{Identifying percolation
phase transitions with unsupervised learning based on largest clusters}}

\author{Dian Xu}
\affiliation{Key Laboratory of Quark and Lepton Physics (MOE) and Institute of Particle Physics, Central China Normal University, Wuhan 430079, China}

\author{Shanshan Wang}
\affiliation{Key Laboratory of Quark and Lepton Physics (MOE) and Institute of Particle Physics, Central China Normal University, Wuhan 430079, China}

\author{Weibing Deng}
\affiliation{Key Laboratory of Quark and Lepton Physics (MOE) and Institute of Particle Physics, Central China Normal University, Wuhan 430079, China}

\author{Feng Gao}
\affiliation{Key Laboratory of Quark and Lepton Physics (MOE) and Institute of Particle Physics, Central China Normal University, Wuhan 430079, China}

\author{Wei Li}
\email[]{liw@mail.ccnu.edu.cn}
\affiliation{Key Laboratory of Quark and Lepton Physics (MOE) and Institute of Particle Physics, Central China Normal University, Wuhan 430079, China}

\author{Jianmin Shen}
\email[]{sjm@mails.ccnu.edu.cn}
\affiliation{Key Laboratory of Quark and Lepton Physics (MOE) and Institute of Particle Physics, Central China Normal University, Wuhan 430079, China}
\affiliation{School of engineering and technology, Baoshan University, Baoshan 678000, China}

\date{\today}

\begin{abstract}
The application of machine learning in the study of phase transitions has achieved remarkable success in both equilibrium and non-equilibrium systems. It is widely recognized that unsupervised learning can retrieve phase transition information through hidden variables. However, using unsupervised methods to identify the critical point of percolation models has remained an intriguing challenge. This paper suggests that, by inputting the largest cluster rather than the original configuration into the learning model, unsupervised learning can indeed predict the critical point of the percolation model. Furthermore, we observe that when the largest cluster configuration is randomly shuffled—altering the positions of occupied sites or bonds—there is no significant difference in the output compared to learning the largest cluster configuration directly. This finding suggests a more general principle: unsupervised learning primarily captures particle density, or more specifically, occupied site density. However, shuffling does impact the formation of the largest cluster, which is directly related to phase transitions. As randomness increases, we observe that the correlation length tends to decrease, providing direct evidence of this relationship. We also propose a method called Fake Finite Size Scaling (FFSS) to calculate the critical value, which improves the accuracy of fitting to a great extent.

\end{abstract}
\maketitle

\section{Introduction}
\label{intro}
Machine Learning (ML)~\cite{jordan2015machine,goodfellow2016machine} is a specialized area of artificial intelligence (AI) that focuses on creating algorithms~\cite{Batta2020machine} and statistical models~\cite{engel2001statistical}. These models enable computers to perform tasks without explicit instructions~\cite{mehta2019high}. ML allows computers to learn from data~\cite{carleo2019machine}, enabling them to make decisions, predictions, or identify patterns independently~\cite{carrasquilla2020machine,ajakan2014domain}. This capability is particularly valuable for data analysis in the field of physics. It offers an alternative approach, alongside theoretical calculations~\cite{hammersley2013monte}, numerical simulations~\cite{domb1996critical,hinrichsen2000non}, and field theory~\cite{amit2005field}, for analyzing phase transitions in statistical physics.

The application of machine learning in phase transition research began with efforts to leverage its powerful data processing and pattern recognition capabilities to identify phase transition behavior in complex systems. Early studies primarily focused on using supervised learning techniques\cite{carrasquilla2017machine,wang2016discovering}, such as neural networks, to identify transition points. By training models on data at known phase transitions, researchers could classify and predict phase transitions for unknown parameters.

In recent years, machine learning approaches have been applied to more complex phase transition problems, including non-equilibrium systems\cite{shen2021supervised}, quantum phase transitions\cite{ohtsuki2016deep}, and topological transitions\cite{deng2017machine,rodriguez2019identifying}. These studies extend beyond simply identifying transition points, exploring instead the microstructure of systems through generative models, such as variational autoencoders (VAEs)\cite{kingma2019introduction,pu2016variational} and generative adversarial networks (GANs)\cite{goodfellow2020generative}. Additionally, research involving graph neural networks (GNNs) has emerged\cite{wu2020comprehensive}, allowing researchers to handle physical systems with complex interactions, further advancing the depth of machine learning applications in the field of phase transitions.\cite{ma2023jet}

With the development of unsupervised learning, numerous studies have shown that techniques such as dimensionality reduction and clustering can automatically capture information about order parameters in phase transition systems. Sebastian J. Wetzel examined unsupervised learning techniques, specifically Principal Component Analysis (PCA)\cite{abdi2010principal} and a neural-network-based Variational Autoencoder (VAE), to identify the features that best describe configurations of the 2D Ising model and the 3D XY model. His findings indicated that the latent parameters corresponded closely with the known order parameters.\cite{wetzel2017unsupervised}

Hu et al applied PCA to studying the phase behavior and transitions in several classical spin models, including the Ising models on square and triangular lattices, and the 2D XY model. They found that the principal components derived through PCA could not only reveal different phases and symmetry breaking but also distinguish between types of phase transitions and locate critical points. Their study also applied autoencoders, demonstrating that these models could be trained to capture phase transitions and identify critical points as well.\cite{hu2017discovering}

Beyond equilibrium phase transition models, Wang et al employed unsupervised learning to analyze the 1+1 dimensional even-offspring branching-annihilating random walks model, an example of nonequilibrium phase transitions, successfully obtaining critical exponents and related information.\cite{wang2024supervised} Further studies on topological\cite{deng2017machine} and quantum phase transitions\cite{dong2019machine} have also demonstrated the effectiveness of unsupervised learning methods in identifying critical characteristics in these systems.

Research on percolation models has a long history, beginning with studies of how fluids diffuse through the pores of coal.\cite{broadbent1957percolation} Modern percolation theory has evolved to focus on changes in network behavior as nodes or edges are added.\cite{kirkpatrick1973percolation} The work in \cite{shante1971introduction} represents one of the earliest systematic discussions of the physical and geometric properties of percolation models, suggesting that the percolating cluster is most likely to emerge within the largest cluster of the model.

With further advancements, percolation models have found increasingly close connections with other fields. For example, Artime et al reviewed the behavior of percolation in cascading failures, providing an overview of the theoretical and computational approaches to robustness and resilience in complex networks.\cite{artime2024robustness} Ji et al also examined the intricate interplay between network structure and signal propagation, contributing to the study of the complex dynamics within interconnected systems.\cite{ji2023signal}

The order parameter in percolation models is not the density of active sites; it also includes the probability that lattice sites (or bonds) belong to the percolating cluster. As a result, using unsupervised learning to identify the critical point in percolation models has been a persistent challenge. Zhang Wanzhou applied an Ising mapping approach to map the original configurations of the percolation model, subsequently using machine learning to identify the system’s critical point.\cite{zhang2019machine} Shu Cheng and colleagues used various neural networks to study configurations with noise.\cite{cheng2021machine} However, none of these studies demonstrated that unsupervised learning alone could directly obtain the critical point from the original configurations of a percolation model.

Jianmin Shen’s unsupervised learning results on the 1+1 dimensional directed percolation (DP) model indicate that both the first principal component from PCA and a single latent neuron from an autoencoder effectively represent the DP model’s order parameter, namely, the particle density. Interestingly, when the lattice configurations are randomized, the unsupervised learning results show no difference from those obtained with the original configurations.\cite{jianmin2023machine} This raises the question: Do the results from unsupervised learning truly represent the order parameter?

Inspired by these findings, in this paper, we propose using a largest-cluster approach combined with the Monte Carlo method to compute active site density, Principal Component Analysis (PCA), and an autoencoder (AE)\cite{ng2011sparse} to locate the critical point in the percolation model. This approach is crucial for examining the relationship between active site density and criticality. Additionally, we investigate the results of unsupervised learning after randomizing the largest cluster. Our main methods include rearranging or selectively modifying the largest cluster before inputting it into algorithms to observe changes in output behavior.

The primary structure of this paper unfolds as follows. Section~\ref{Models} elucidates the percolation model configurations of interest. Section~\ref{method} expounds upon the two methodologies of unsupervised Learning. Section~\ref{usml_result} delineates the research findings, wherein a comparative analysis of six distinct configurations under MC, PCA, and AE is presented. Finally, in Section~\ref{Conclusion}, we furnish a comprehensive summary of this study.

\section{Model and unsupervised Learning}
\subsection{percolation model}\label{Models}
\begin{figure*}[htbp]
\begin{tabular}{cc}
    \includegraphics[width=0.35\textwidth]{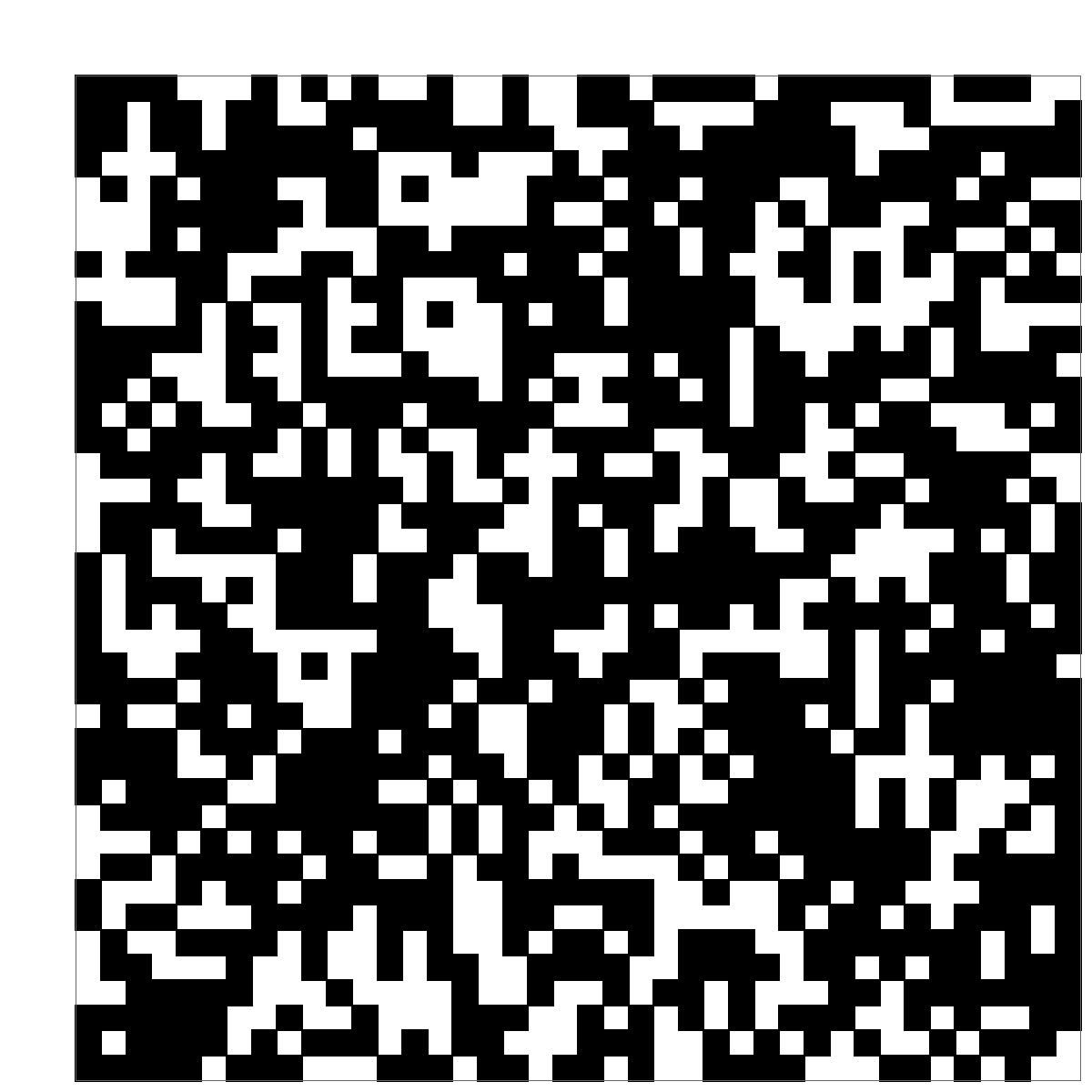} &
    $\qquad$\includegraphics[width=0.35\textwidth]{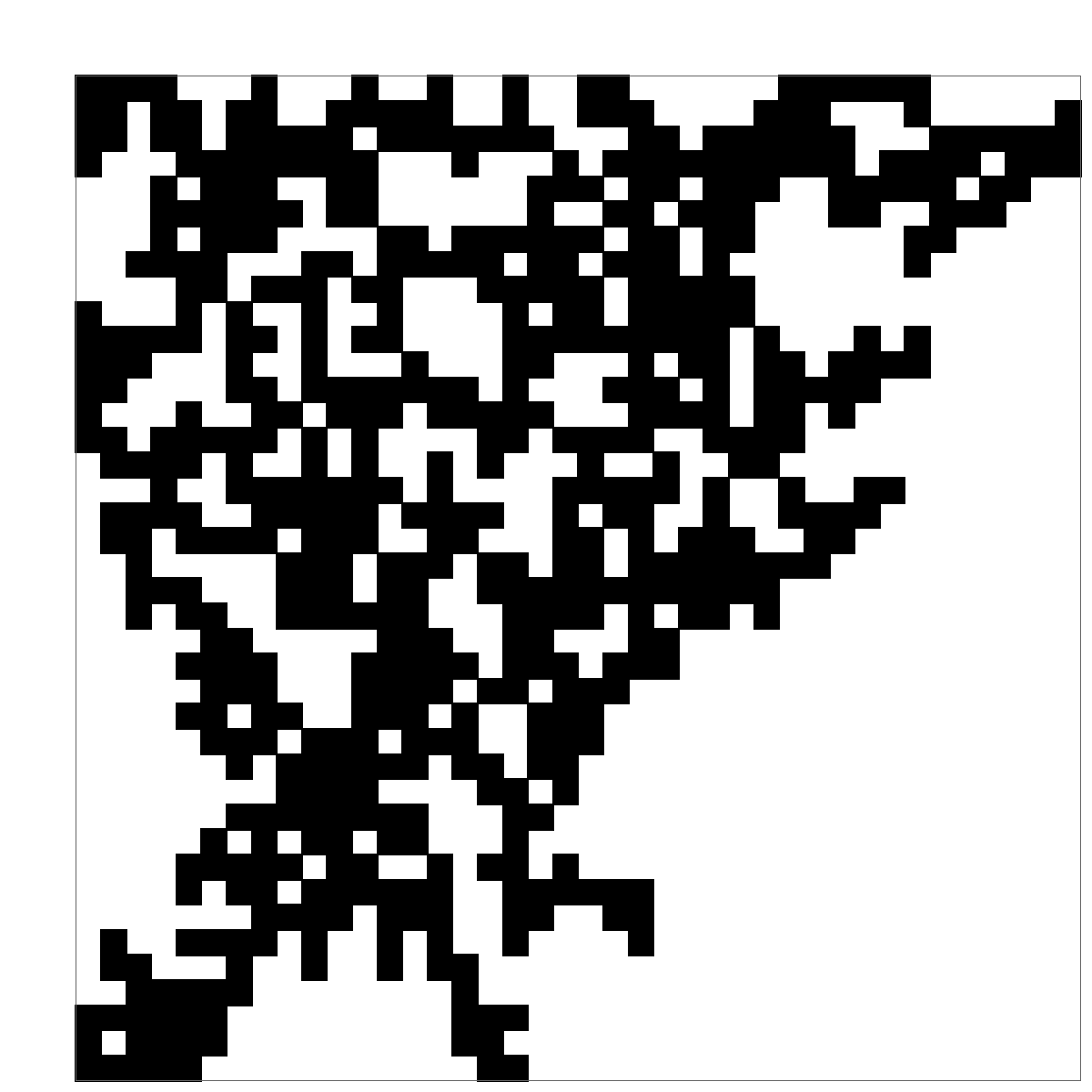} \\
    (a) & $\qquad$ (b)\\
    \includegraphics[width=0.35\textwidth]{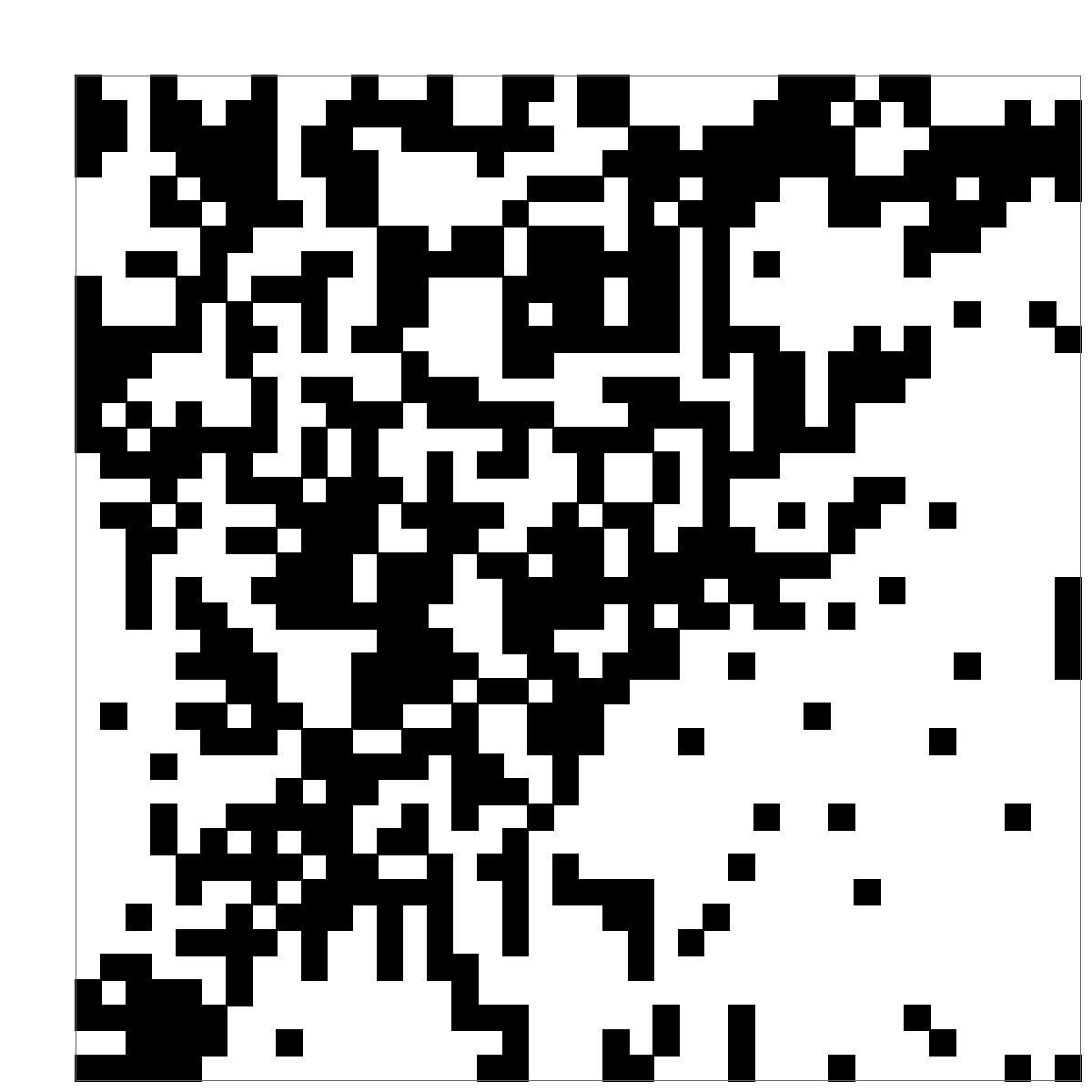} &
    $\qquad$ \includegraphics[width=0.35\textwidth]{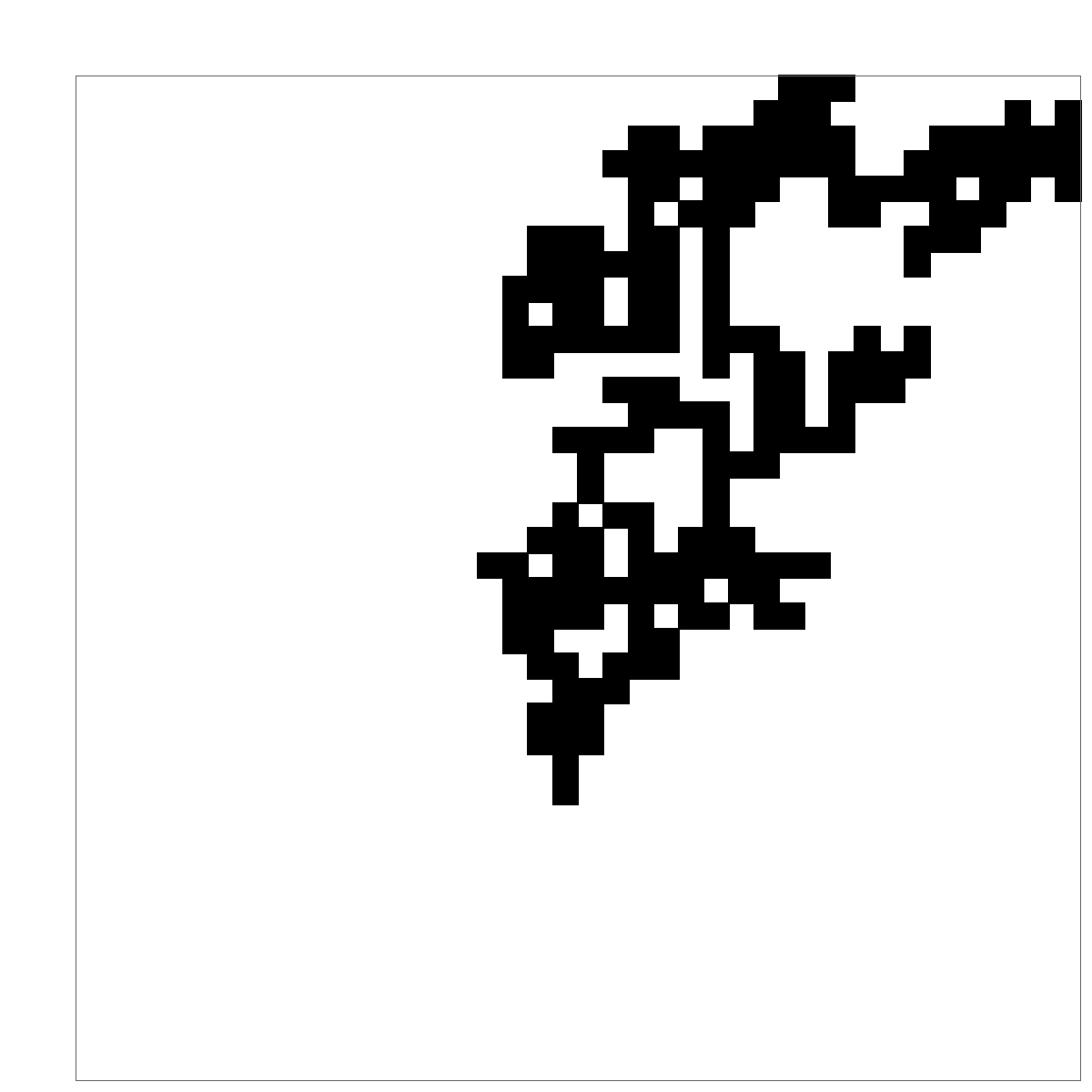} \\
    (c) & $\qquad$(d)\\
     \includegraphics[width=0.45\textwidth]{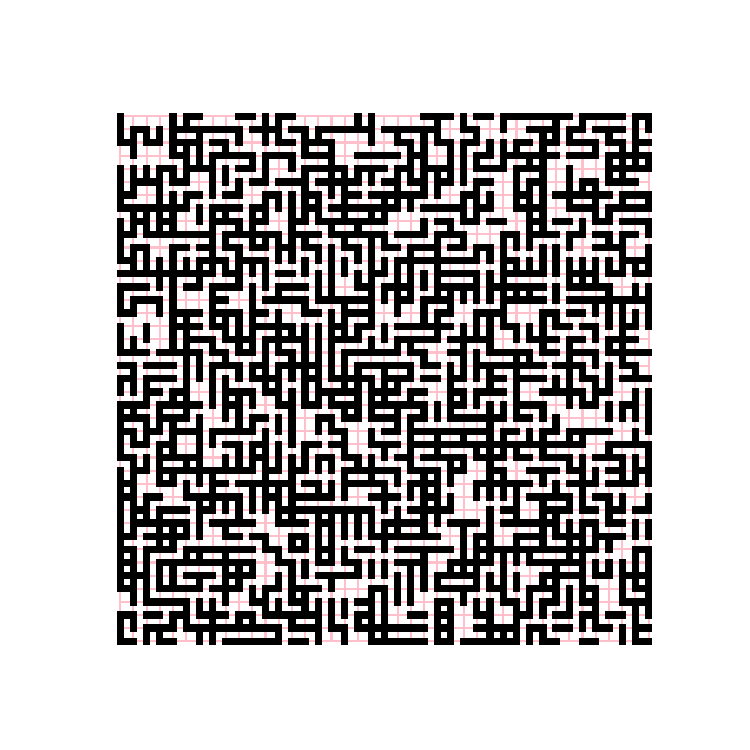} &
    $\qquad$ \includegraphics[width=0.45\textwidth]{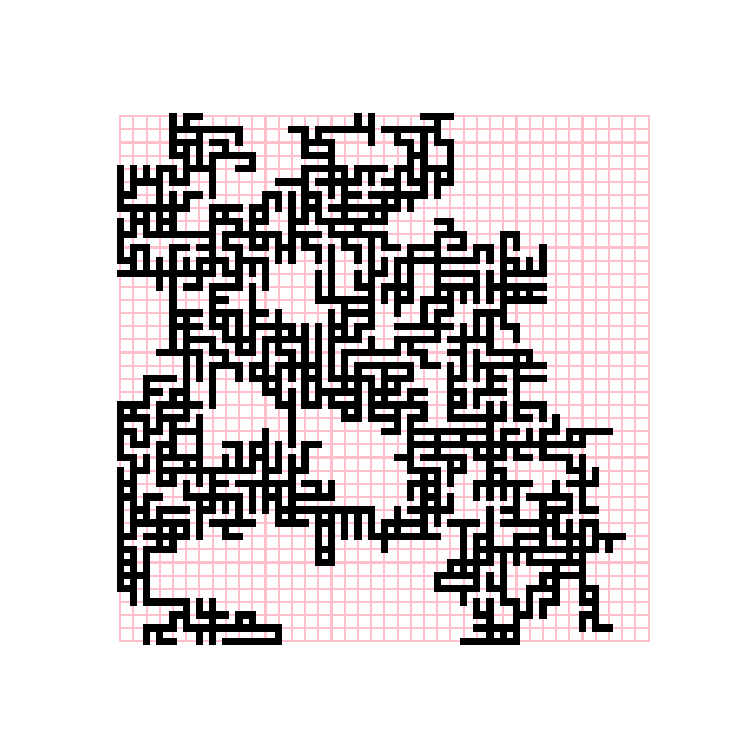} \\
    (e) & $\qquad$(f)
\end{tabular}
\caption{In this article, we examine six distinct configurations. \textbf{a} is the raw configuration of site percolation with occupation probability = 0.8. \textbf{b} is the largest cluster of \textbf{a}. \textbf{c} shows the shuffled configuration of \textbf{b} with a ratio $r = 0.2$ . \textbf{d} shows the largest cluster of figure \textbf{c}. \textbf{e} is the raw configuration of bond percolation with occupation probability = 0.8. \textbf{f} shows the largest cluster of \textbf{e}.}
\label{percolation_configuration}
\end{figure*}

The two-dimensional percolation model represents a continuous phase transition,\cite{riordan2011explosive} with the order parameter given by
\begin{equation}
   P_{\infty} (p) \propto (p - p_{c})^{\beta}  \qquad for \quad p \rightarrow p_{c} ^{+}
\end{equation}
where $p$ is the occupation probability, $p_{c}$ is the critical probability,commonly referred to as the percolation threshold. $\beta$ is the critical exponent of the order parameter, and $P_{\infty} (p)$ denotes the probability that a given site or bond belongs to the percolating cluster namely percolation probability. Typically, numerical simulation methods are employed to compute $P_{\infty} (p)$ to locate the critical point.

In the context of a square lattice, the probability of encountering a percolating cluster within the system remains extremely low as long as the probability remains below the critical probability $p_c$. Conversely, when the probability surpasses the critical probability $p_c$, the likelihood of encountering a percolating cluster rapidly approaches unity as the number of occupied sites increases.

The percolating cluster, which is the focus of this study, contains the critical information of the percolation phase transition. In percolation models, the raw configurations include not only the percolating cluster but also other isolated sites or bonds. 

In finite systems, the size of the largest cluster, $S_{\text{max}}$, serves as a crucial observable for understanding critical behavior and can be expressed as:
\begin{equation}
    S_{\text{max}} \sim L^{d-\beta/\nu} \cdot f\left((p - p_c)L^{1/\nu}\right),
\end{equation}
where L is linear size of the system, with $L^d$ representing the total number of sites in the system. $\nu$ and $\beta$ are  critical exponents, while $f(x)$ is a universal scaling function. Specifically, as $p \to p_c^-$, the size of the largest cluster grows rapidly but remains finite.
As $p \to p_c^+$, the largest cluster becomes the infinite percolation cluster, with its occupancy fraction $P_{\infty} (p)$ obeying the scaling relationship dictated by the critical exponent $\beta$.

Typically, the phenomenon of percolation manifests predominantly within the largest cluster. Consequently, this article abstains from considering secondary or smaller clusters. For two-dimensional site percolation, the critical value is $p_{c} = 0.592746$,the bond percolation one is $p_{c} = 0.5$.~\cite{christensen2005complexity}.The largest clusters are extracted using the depth-first search algorithm.

To probe the percolation model , Commence by constructing a lattice comprising $N \times N$ sites, and adjacent sites are connected by bonds . For site percolation we set every bond have been occupied then progress to randomly populate each site with a specified probability, designated as $p$ , for the bond percolation model, operate the opposite.  Assign labels to the connected clusters of occupied sites or bond. Two occupied one are deemed part of the same cluster if they are contiguous to each other. Scrutinize whether any of the clusters traverse the lattice from one end to the other, signifying percolation. To procure reliable statistics, iterate through the steps mentioned earlier multiple times.

\subsection{Unsupervised Learning}\label{method}
\subsubsection{PCA}
A primary application of PCA lies in diminishing the data's dimensionality while conserving maximal variability. This proves advantageous for data visualization, feature reduction, or preparing the data for subsequent machine learning algorithms. PCA entails the computation of the eigenvalues and eigenvectors of the data's covariance matrix. The eigenvectors, denoted as principal components, delineate the orientations of the new feature space, while the eigenvalues signify their magnitude or the variance along those orientations. Subsequent to extracting the principal components, the data can be projected onto them to effectuate dimensionality reduction.

In this paper, we employ the PCA algorithm for dimensionality reduction, which involves several key steps. First, we prepare the input matrix,
\begin{equation}
    X = 
    \begin{bmatrix}
    x_{11} & \cdots & x_{1n} \\
    \vdots & \ddot & \vdots  \\
    x_{m1} & \cdots & x_{mn}
    \end{bmatrix}
\end{equation}
here, $m$ represents the number of occupancy probabilities we consider, denoted as $p_{number}$. In this study, we sample 41 points for the occupancy probability, ranging from 0 to 1 with intervals of 0.02. The size of $n$ corresponds to the product of the system’s length and width. Since we use a square lattice, $n = L * L$. Next, we subtract the mean value of each column from every element in that column, effectively centering the data for each feature.
\begin{equation}
    \bar x_{a} = \frac{x_{a1} + x_{a2} + \cdots + x_{an}}{n}
\end{equation}

After preprocessing, the data is standardized to have a mean of 0 and a standard deviation of 1. We define this new matrix as $Y$. Subsequently, we compute the covariance matrix of $Y$, which captures the relationships between the features in the standardized data.
\begin{equation}
    A = \frac{1}{n}YY^T
\end{equation}

Generally, there are two methods to compute the eigenvalues and eigenvectors of the covariance matrix. In this paper, we use the approach based on Singular Value Decomposition (SVD) of the covariance matrix. For any matrix $A$, an SVD always exists, which can be expressed as:
\begin{equation}
    A = U \Sigma V^T
\end{equation}
where $U$ and $V$ are orthogonal matrices, and $\Sigma$ is a diagonal matrix containing the singular values of $A$.By applying SVD to the covariance matrix, we can efficiently obtain its eigenvalues and eigenvectors, which are essential for performing PCA.

Based on this understanding, we proceed as follows:Firstly,We calculate the eigenvalues and eigenvectors of $AA^T$. After normalizing the eigenvectors, they form the matrix $U$.Similarly, we calculate the eigenvalues and eigenvectors of  $A^T A$. After normalizing the eigenvectors, they form the matrix $V$. The eigenvalues obtained from $AA^T$  are square-rooted to form the singular values. These singular values are arranged along the diagonal of the matrix $\Sigma$.where  $U$ ,  $\Sigma$ , and  $V$  correspond to the left singular vectors, singular values, and right singular vectors, respectively.

To compress the data into a single dimension, from the diagonal matrix  $\Sigma$ , we select the largest singular value. Squaring this value gives us the largest eigenvalue of $AA^T$, which is also the largest eigenvalue of $A^T A$ .We extract the eigenvector of $A^T A$ that corresponds to this largest eigenvalue. This eigenvector forms a row vector, which we transpose to create a new matrix $K$ .The new matrix  $K$  is then multiplied by the original input matrix X 
\begin{equation}
    pca_1 = X^{'}_{m*1} = X_{m*n} K_{n*1}
\end{equation}
the resulting matrix $X^{'}_{m*1}$ is the desired one-dimensional compressed data, referred to as $pca_1$. Finally, by averaging the values across 1000 samples, we obtain the final dataset $X^{''}_{p_{number}*1}$. This step consolidates the compressed data for each occupancy probability, providing a concise and meaningful representation for further analysis.

\subsubsection{AE}
\begin{figure*}[htbp]
\centering
\includegraphics[width=0.8
\textwidth]{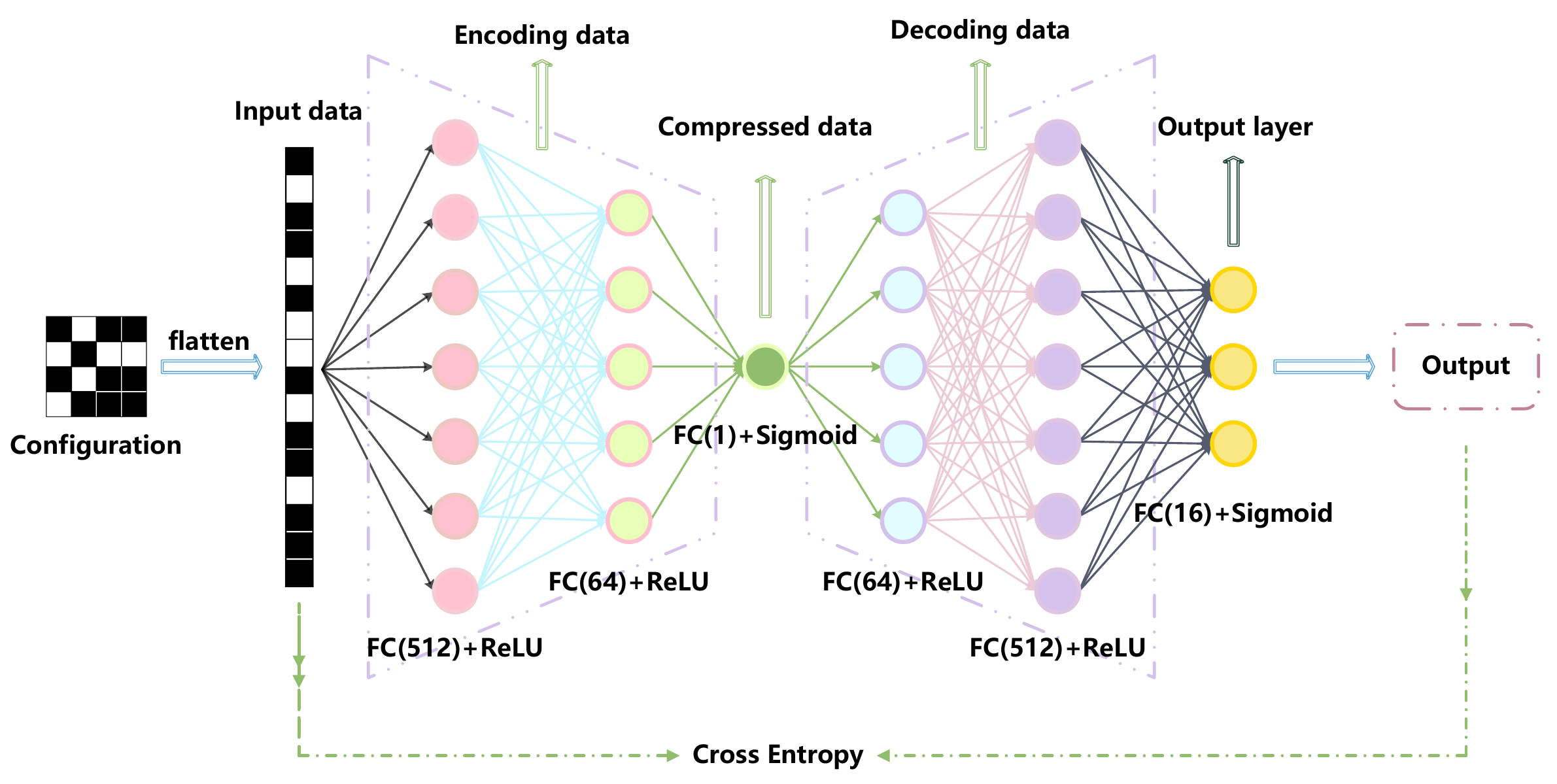}
\caption{Neural network schematic structure of autoencoder.}
\label{fig:ae}
\end{figure*}

An Autoencoder (AE) is a type of artificial neural network architecture in the field of unsupervised learning, consisting of two key components: the encoder and the decoder.

The encoder compresses the input data into a more compact representation, aiming to capture the intrinsic structure of the original dataset. This serves a dual purpose: reducing the dimensionality of the data and minimizing the impact of noise.

In contrast, the decoder reconstructs the input data from its compressed representation, striving to generate an output that faithfully reflects the original input.

The workflow of an Autoencoder is illustrated in the FIG \ref{fig:ae}. 

In the Autoencoder (AE) workflow presented in the FIG \ref{fig:ae},  we take the case of $L=4$ as an example. The process begins by flattening the original configuration. Specifically, the elements of the $L*L$ matrix are concatenated row by row: the first row is appended to the start of the second row, then the new second row is appended to the third row, and so on. Ultimately, we obtain a $1*L^{2}$ matrix representation denoted as $R$.

Next, this flattened matrix $R$ is fed into a layer with 512 neurons. Whether each neuron is activated depends on the following transformation process:
\begin{equation}
    h_i=f(\sum\limits_{j}w_{ij}R_{j}+b_{i})
\end{equation}
where $w_{ij}$ represents the weights connecting input $j$ to neuron $i$, $b_{i}$ is the bias term associated with neuron $i$,
$f()$ is the activation function (e.g., ReLU or sigmoid), and
$h_i$ is the output of neuron $i$.

This transformation reduces the original high-dimensional input to a compact, encoded representation. The encoding layer captures the critical features of the data, which are later used by the decoder to reconstruct the original configuration.

Through layer-by-layer propagation, we eventually obtain a matrix $T$ that is identical in form to the original matrix $R$, with dimensions $1*L^{2}$. The network weights are updated using gradient descent via backpropagation. In this process, the error (or loss) between the predicted output $T$ and the target input $R$ is propagated back through the network to adjust the weights in order to minimize the loss.

The weight updates are based on the gradient of the loss function with respect to the network parameters. Specifically, the update rule is:
\begin{align}
    W \leftarrow W-\eta\frac{\partial L}{\partial W}  \\
    b \leftarrow b-\eta\frac{\partial L}{\partial b}
\end{align}
where W and b represent the weight matrix and bias terms of the network, $\eta$is the learning rate, $\frac{\partial L}{\partial W}$and $\frac{\partial L}{\partial b}$ are the gradients of the loss function $L$ with respect to the weights and biases.

Through this iterative process of adjusting the weights, the neural network learns to approximate the original configuration and reduce the difference between $R$ and $T$. This enables the network to effectively capture the underlying structure of the data, facilitating tasks such as reconstruction, feature extraction, or anomaly detection in the context of unsupervised learning.

The ultimate goal is to minimize the cross-entropy loss function between $R$ and $T$, which will lead to the network’s output. Cross-entropy can be understood as the measure of difficulty in representing the probability distribution $R$ using the probability distribution $T$. Its expression is given by:
\begin{equation}
    H(R,T) = \sum R(x_i) log \frac{1}{T(x_i)}
\end{equation}
Where $R(x_i)$ is the true probability distribution (the original data), $T(x_i)$ is the predicted probability distribution (the output of the neural network), The summation runs over all elements $x_i$ in the data.

The objective of the network is to adjust its parameters in such a way that the predicted distribution $T$ gets as close as possible to the true distribution $R$, thereby minimizing the cross-entropy loss. By minimizing this loss, the network learns the underlying structure of the data, enabling effective data reconstruction or feature representation.

We extract the output values $h$ from the hidden layer of the optimized network. By averaging the values for all samples under the same probability, we can obtain the desired research object. This process allows us to analyze the underlying features learned by the network, providing insights into the system’s characteristics, such as phase transitions or other phenomena related to the percolation model or other complex systems being studied.
 
\section{The Unspervised Learning Results}
\label{usml_result}

Typically, the first principal component from Principal Component Analysis (PCA) and the single latent variable from the Autoencoder (AE) are considered crucial in extracting key values from phase transition models. To clarify this relationship, we investigate several different configurations of the percolation model, namely the raw configuration, the largest cluster, the shuffled largest cluster, and the shuffled largest cluster with rearranged sites, as shown in Figure \ref{percolation_configuration}.

To facilitate comparison and analyze the accuracy of the results, we first perform a Monte Carlo (MC) simulation on the raw configuration. The maximum derivative value of the function is considered as the critical point of the model. It is worth noting that the results presented in this paper have been normalized to ensure consistency.

\begin{figure*}[!thb]
\begin{tabular}{ccc} 
    \includegraphics[width=0.45\textwidth]{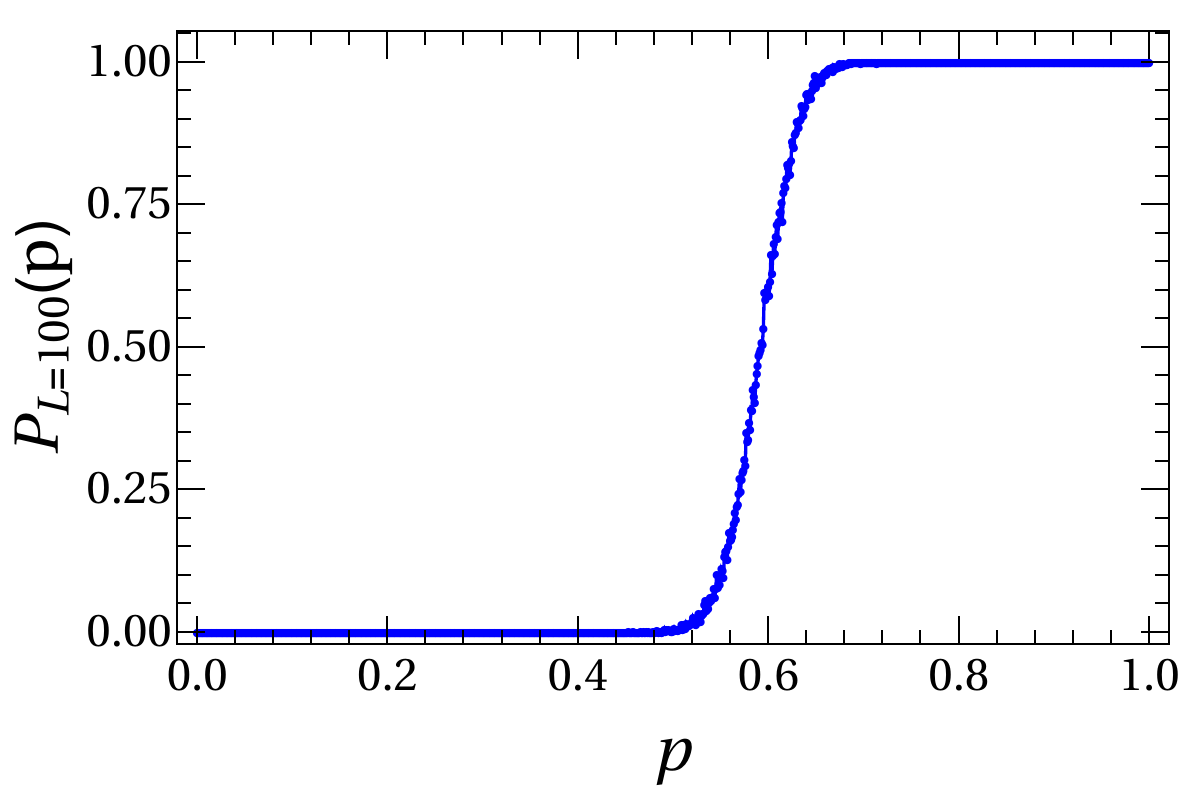}&
    \includegraphics[width=0.45\textwidth]{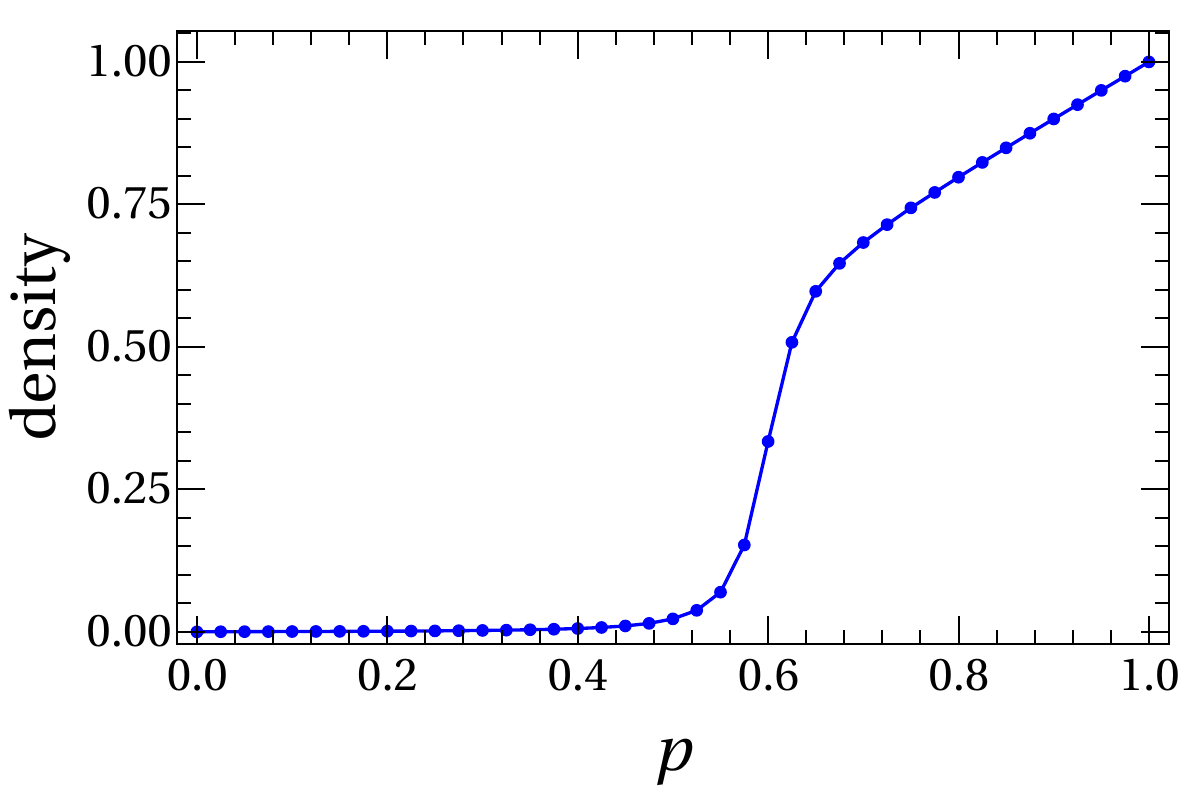}\\
    (a) & (b)  \\
    \includegraphics[width=0.45\textwidth]{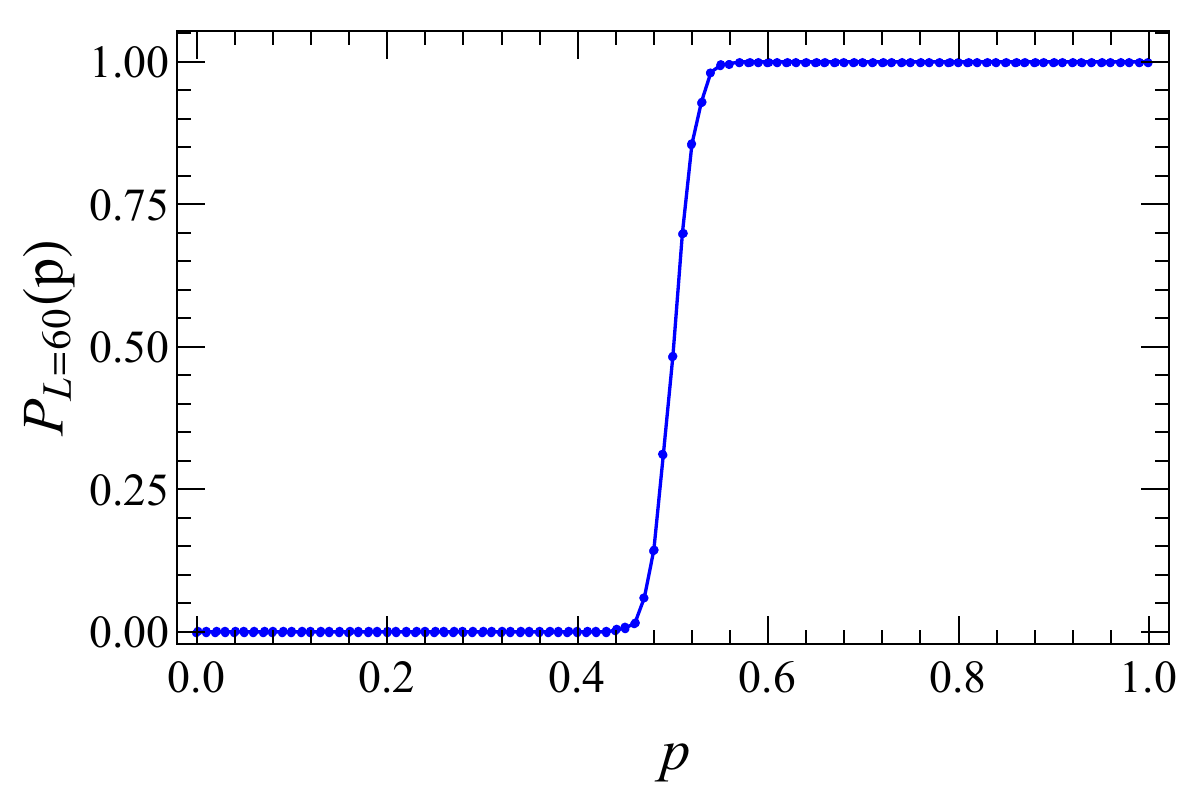}&
    \includegraphics[width=0.45\textwidth]{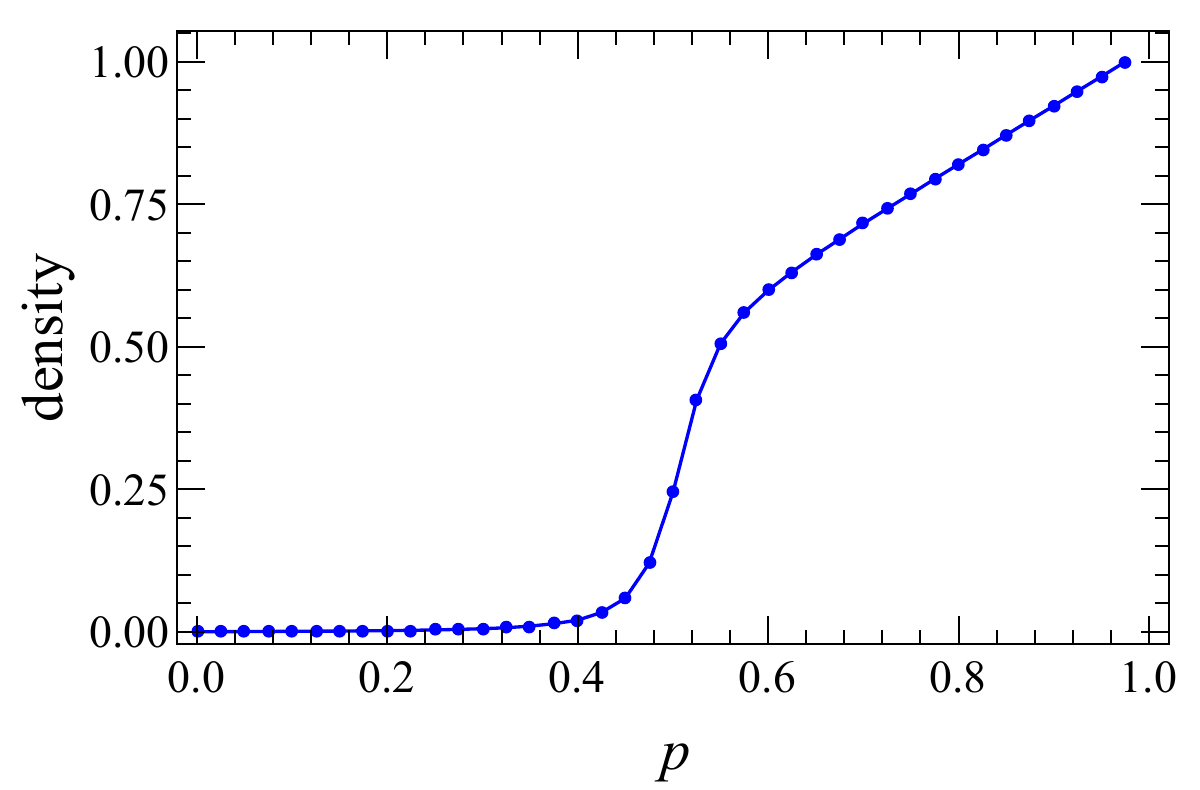}\\
    (c) & (d)
\end{tabular}
\caption{The MC simulation results about site and bond percolation of a two-dimensional system of size $L \times L$. (a) means the probability that a site belongs to a percolating cluster, (b) calculated the largest cluster's density of active sites of the system.(c) means the probability that a bond belongs to a percolating cluster, (d) calculated the largest cluster's density of active bonds of the system.}
\label{percolation order}
\end{figure*}

As shown in Figures \ref{percolation order}(a) and \ref{percolation order}(c), we conducted site percolation simulations on a lattice of size $L=100$ with 1001 occupation probability values ranging from 0 to 1, and bond percolation simulations on a lattice of size $L=60$ with 41 occupation probability values within the same range. For each probability, 1000 lattice configurations were generated, and the number of percolation occurrences was recorded.

The $y$-axis values are obtained by dividing the percolation counts by 1000, while the $x$-axis represents the generation probabilities. Notably, Figure \ref{percolation order}(a) shows a sharp transition peak near 0.59, which aligns closely with the theoretical value of 0.593, as expected. Similarly, Figure \ref{percolation order}(c) reveals a transition near 0.5, consistent with the expected critical threshold for bond percolation.

On the other hand, we investigated the correlation between the density of active sites/bonds(see equation\ref{eqdensity}) of the largest cluster and the critical points in the percolation models. Here, the active density is defined as the ratio of occupied sites/bonds to the total number of lattices. We extracted the largest cluster configurations from the site and bond percolation models described above and calculated the density of active sites/bonds. The average of these values constitutes the $y$-axis of the plot, while the $x$-axis represents the generation probabilities.
\begin{equation}
    \rho=\frac{\sum{occupied\ sites/bonds}}{total\ sites/bonds}
    \label{eqdensity}
\end{equation}

As shown in Figure~\ref{percolation order}(b), the curve exhibits a notable rapid change near the critical point, closely aligning with the expected value around $0.59$. Similarly, in Figure~\ref{percolation order}(d), a clear transition is observed near $0.5$ as the generation probability increases, consistent with the critical threshold for bond percolation. These results validate the effectiveness of our simulation method based on the largest cluster and provide a solid foundation for the subsequent analyses in this study.

At the same time, we calculate the Pearson correlation coefficient to compare the MC simulation of density of active sites/bonds, the first principal component of the PCA learning result denoted by $pca_1$, and the single hidden variable denoted by $h$ in the AE. The coefficient quantifies the degree of the linear correlation between any pair of variables.
\begin{equation}
r = \dfrac{\sum \limits_{i = 1}^{n} (\rho_{i} - \overline{\rho}) (h_{i} - \overline{h})}{\sqrt{\sum \limits_{i = 1}^{n} (\rho_{i} - \overline{\rho})^{2}} \sqrt{\sum \limits_{i = 1}^{n} (h_{i} - \overline{h})^{2}}},
\end{equation}
\noindent
where $\rho_i$ and $h_i$ are the density of active sites/bonds of system and the single hidden variable at $i$-th site or bond probability $p$, while $\overline{\rho}$ and $\overline{h}$ are the mean values of $\rho_i$ and $h_i$, respectively. In PCA, $h$ is replaced with $pca_1$.

\subsection{Results of percolation}

\subsubsection{The raw configurations}

\begin{figure*}[t]
\begin{tabular}{cccc}
    \includegraphics[width=0.30\textwidth]{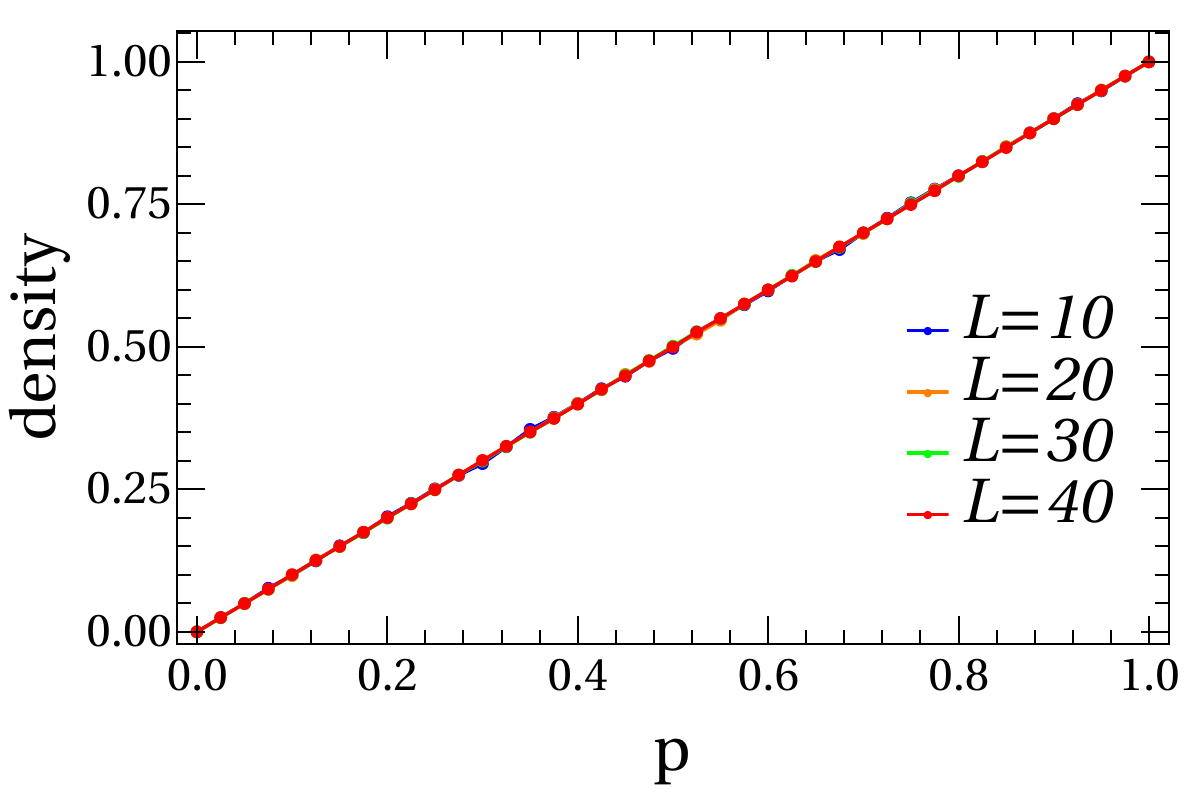} &
    \includegraphics[width=0.30\textwidth]{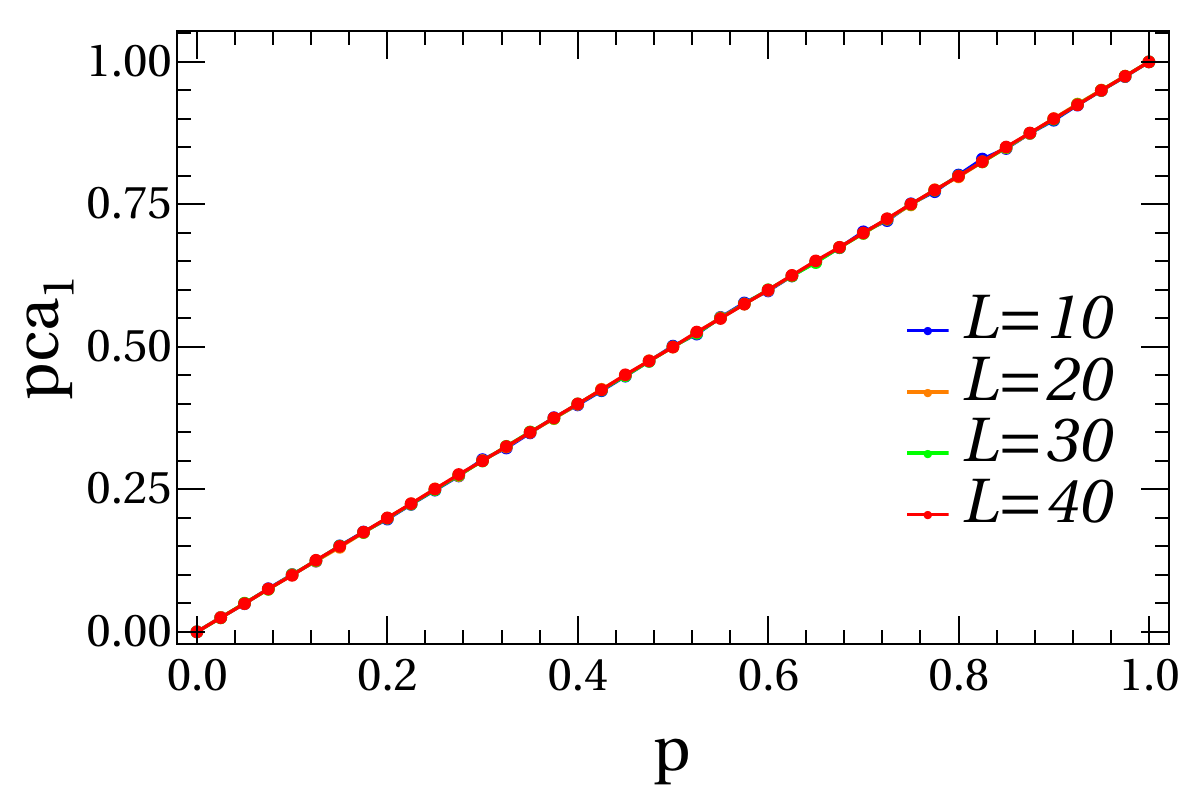}&
    \includegraphics[width=0.30\textwidth]{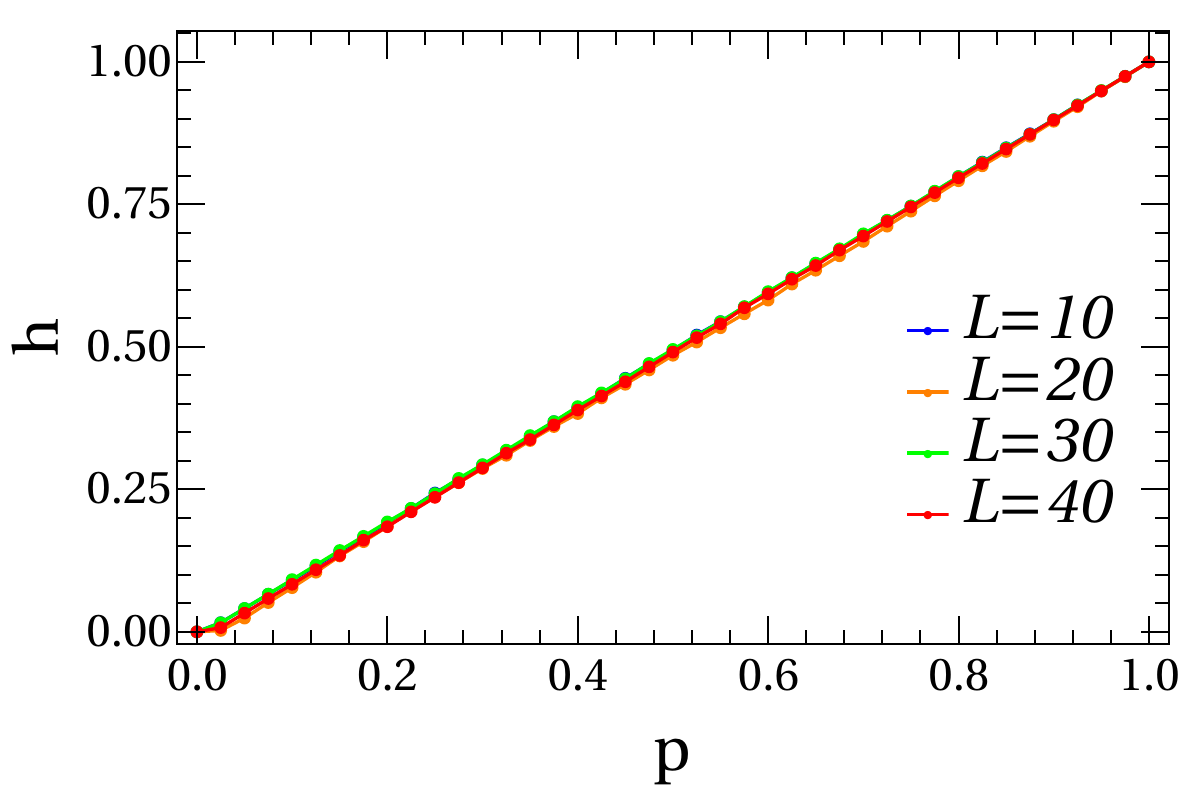} \\
     (a) &  (b) &  (c)   \\
     \includegraphics[width=0.30\textwidth]{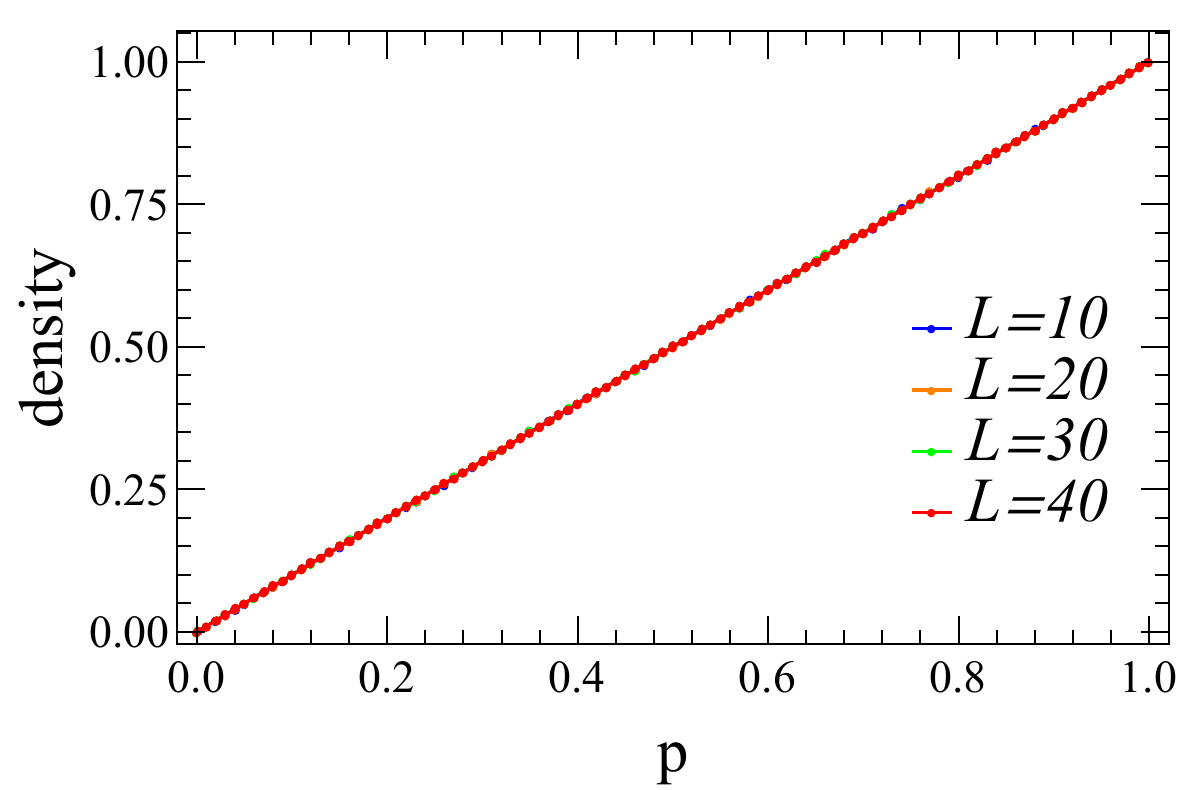} &
    \includegraphics[width=0.30\textwidth]{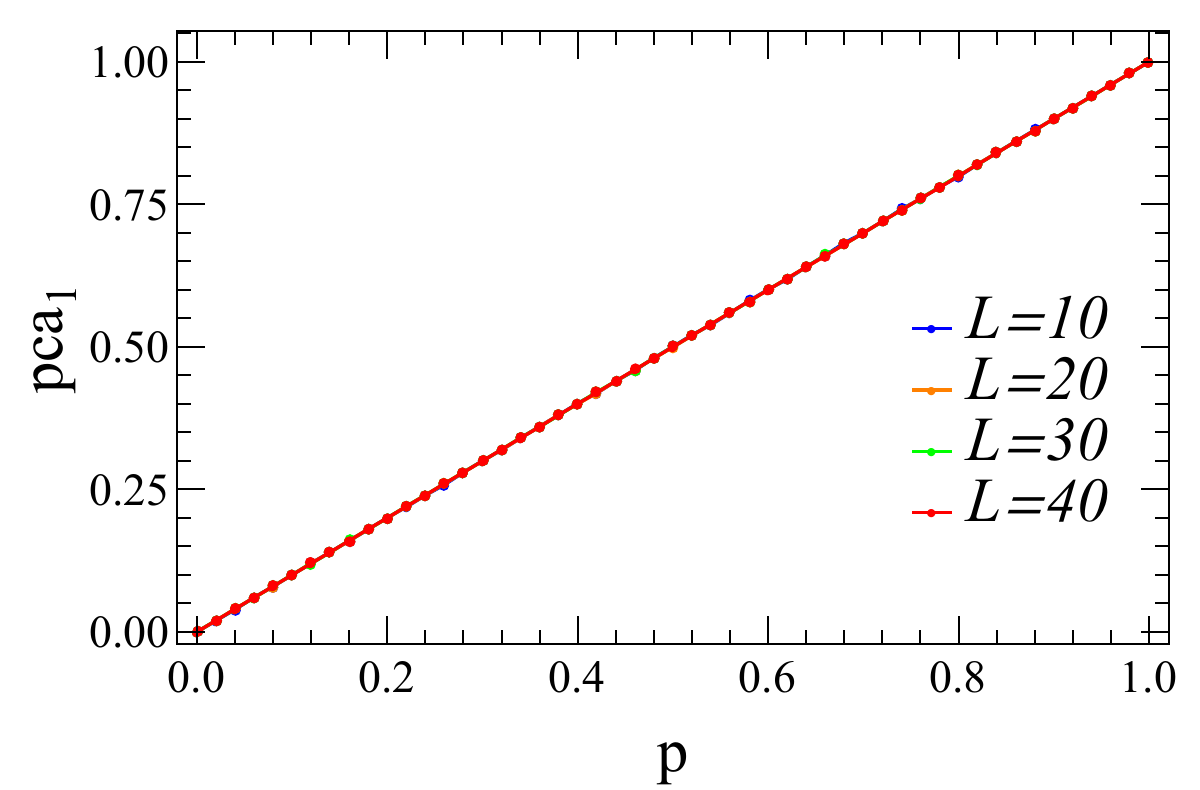}&
    \includegraphics[width=0.30\textwidth]{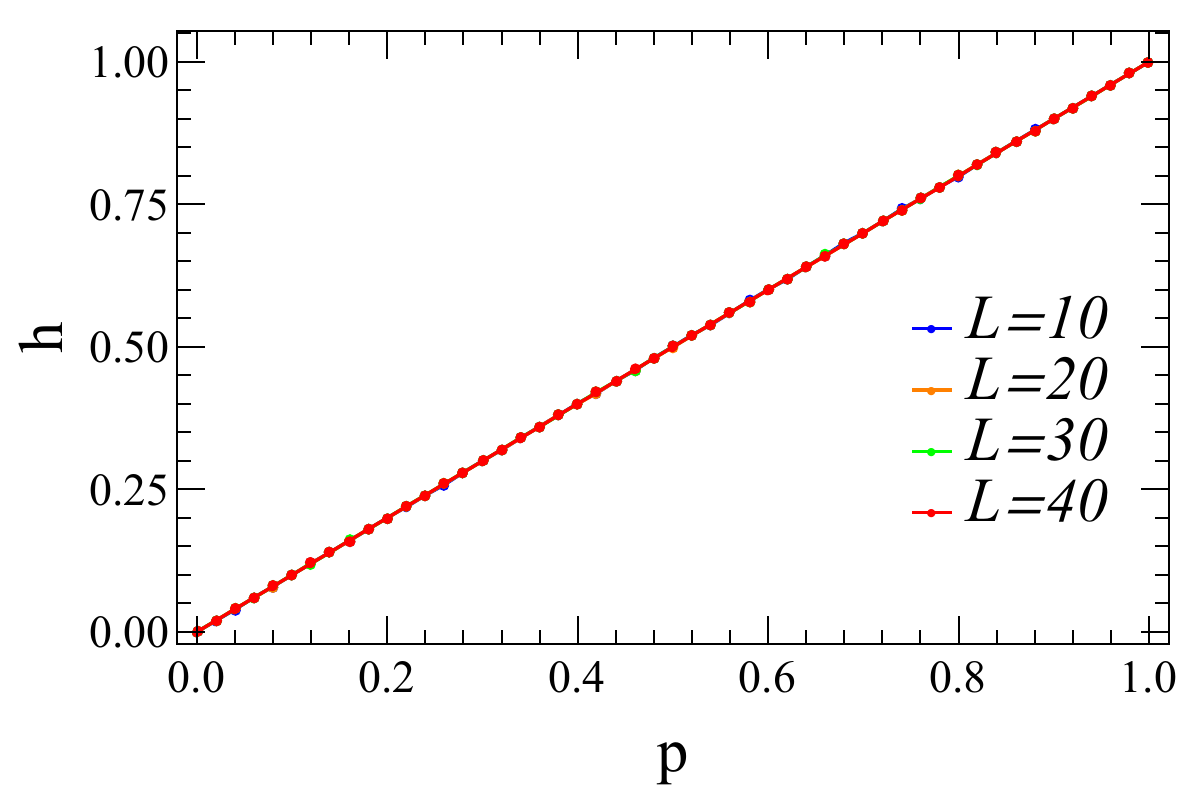}\\
    (d) &  (e) &  (f)
\end{tabular}
\caption{The results of two-dimensional site and bond percolation with the raw configuration. The vertical coordinates of Panels \textbf{a \& d} denote the density of active sites/bonds, \textbf{b \& e}, the first principal component of PCA, and \textbf{c \& f} the single latent variable of AE, respectively. Their horizontal coordinates are all occupation probability. They all exhibit a simple linear increase in nature and overlap, making it impossible to identify the critical point of the system. However, there is a highly similar behavior among the three.}
\label{raw}
\end{figure*}

As an initial baseline, we employed the MC to simulate site and bond percolation model with sizes $L= 10$,$20$,$30$, and $40$, then analyzed the raw configurations. The results, depicted in FIG.~\ref{raw}(a)(d), reveal a linear increase in density of active sites/bonds as the generation probability escalates. In the normalized outcome, densities function forms from various sizes exhibit overlapping behavior. However, the outcome lacks discernible non-trivial features and fails to pinpoint the critical points.

Drawing from our understanding of the percolation model, an augmentation in the generation probability invariably culminates in a linear escalation of the model's density of active sites/bonds, corroborated by the findings of the MC simulation. Nevertheless, what causes our interest is that both PCA, depicted in FIG.~\ref{raw}(b)(e), and AE, depicted in FIG.~\ref{raw}(c)(f), manifest some form of linear relationship. This prompts us to speculate whether the input of the model's configuration information into PCA and AE yields a relationship between the first principal component of PCA ($pca_{1}$) and the single hidden variable ($h$) with respect to density of active sites/bonds. We intend to undertake a verification of this hypothesis in the ensuing chapter.

\subsubsection{The largest cluster}\label{The largest cluster}
\begin{table*}[tbh!]
	\centering
	\begin{tabular}{|c|c|c|c|c|c|}
        \hline
       \diagbox[width=10em]{size}{intersection}{size}     
             &$L_{10}$   &$L_{20}$    &$L_{30}$  &$L_{40}$   \\
        \hline
   $L_{20}$  &$P_{10-20}$&            &          &            \\
        \hline
   $L_{30}$  &           &$P_{20-30}$ &          &           \\
        \hline
   $L_{40}$  &           &            &$P_{30-40}$&          \\
        \hline
   $L_{50}$  &           &            &           &$P_{40-50}$ \\
        \hline
   \end{tabular}
\caption{A schematic diagram of the Fake Finite Size Scaling (FFSS) method, used to identify the intersection points of function curves obtained from experimental results for different system sizes. In the table, both the rows and columns represent function curves of the model at varying system sizes.}
\label{FFSS}
\end{table*}

\begin{figure*}[t]
\begin{tabular}{cccc}
    \includegraphics[width=0.30\textwidth]{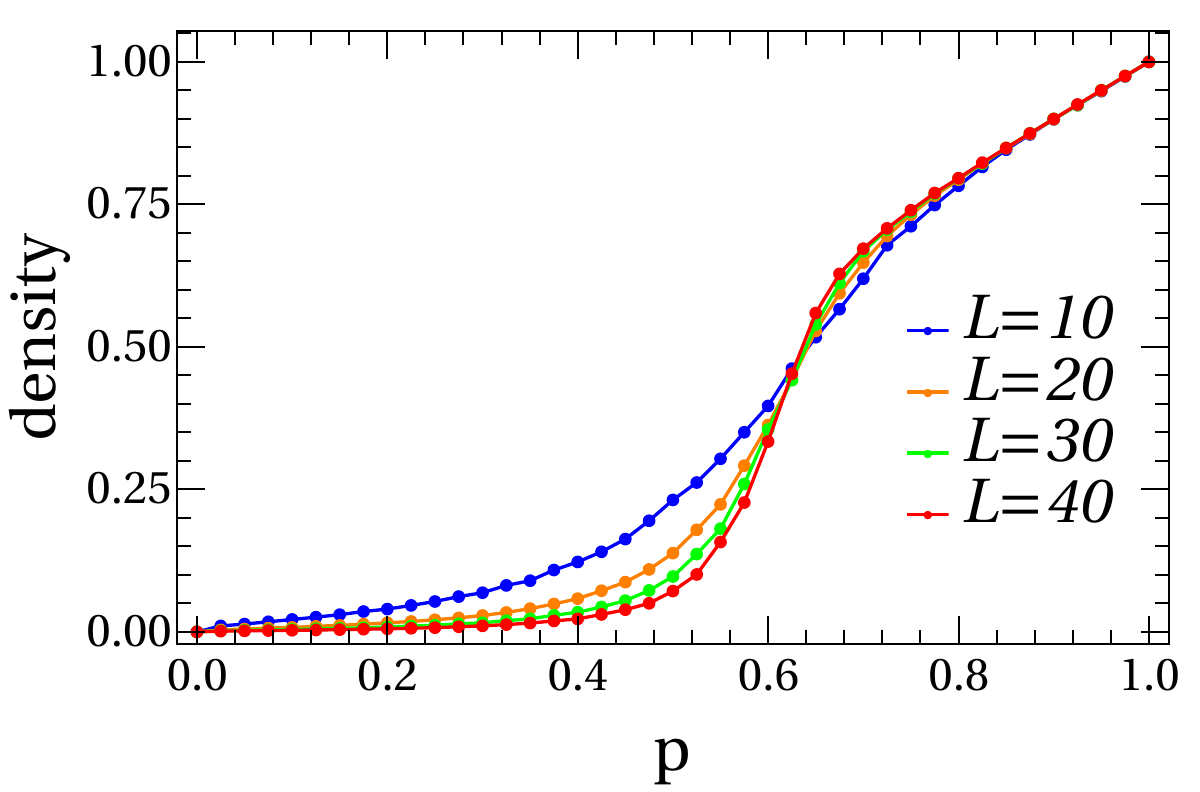}&
    \includegraphics[width=0.30\textwidth]{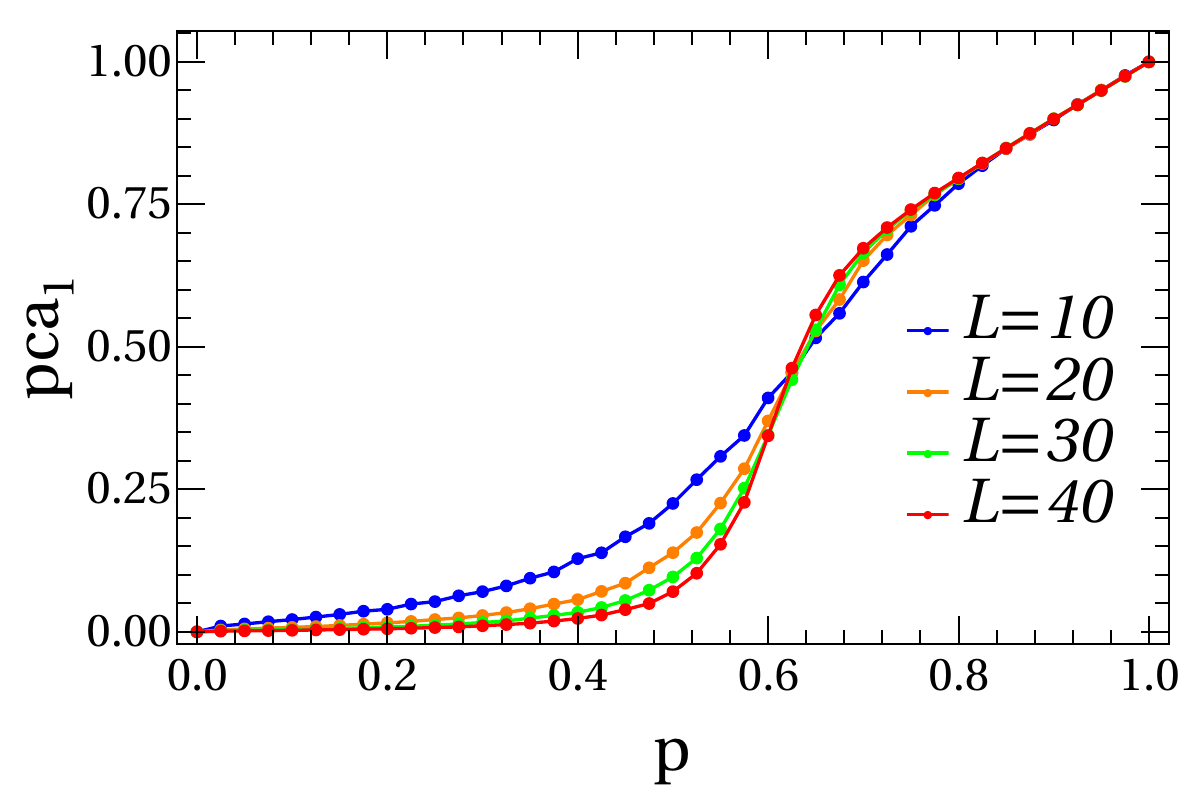}& \includegraphics[width=0.30\textwidth]{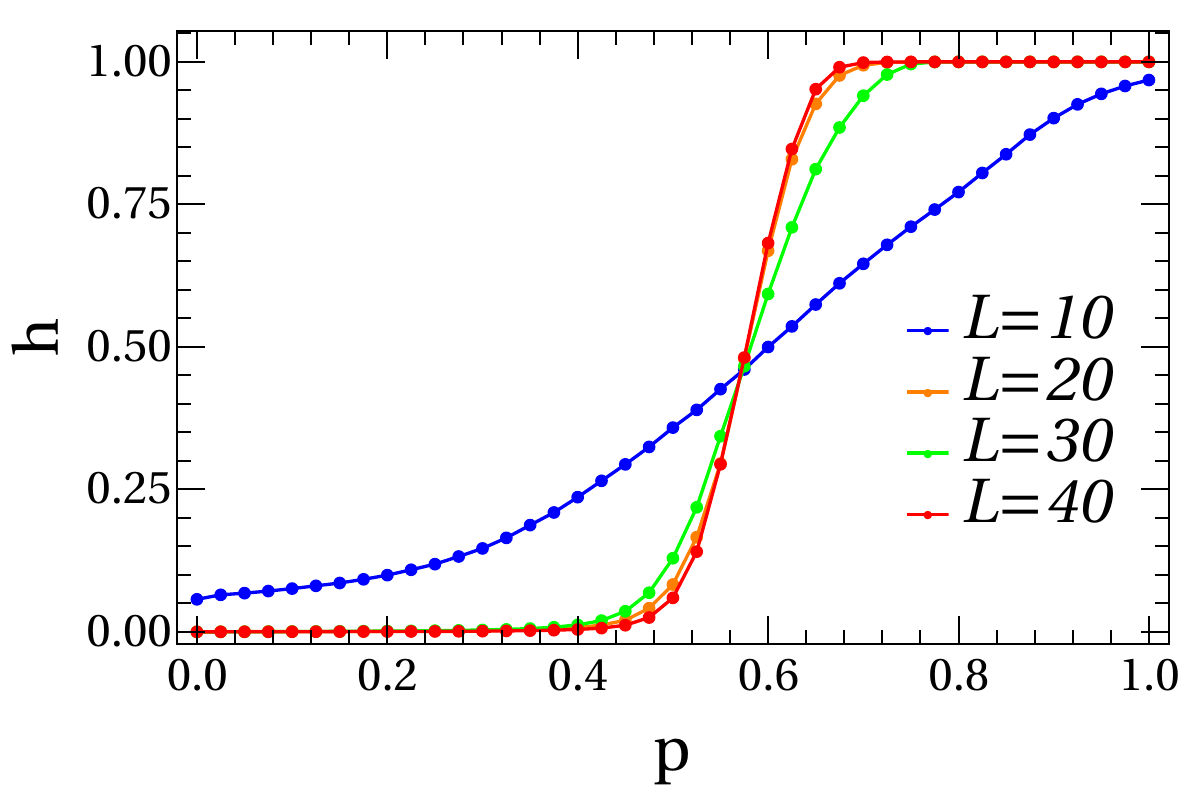} \\
    (a) &  (b) & (c)   \\
    \includegraphics[width=0.30\textwidth]{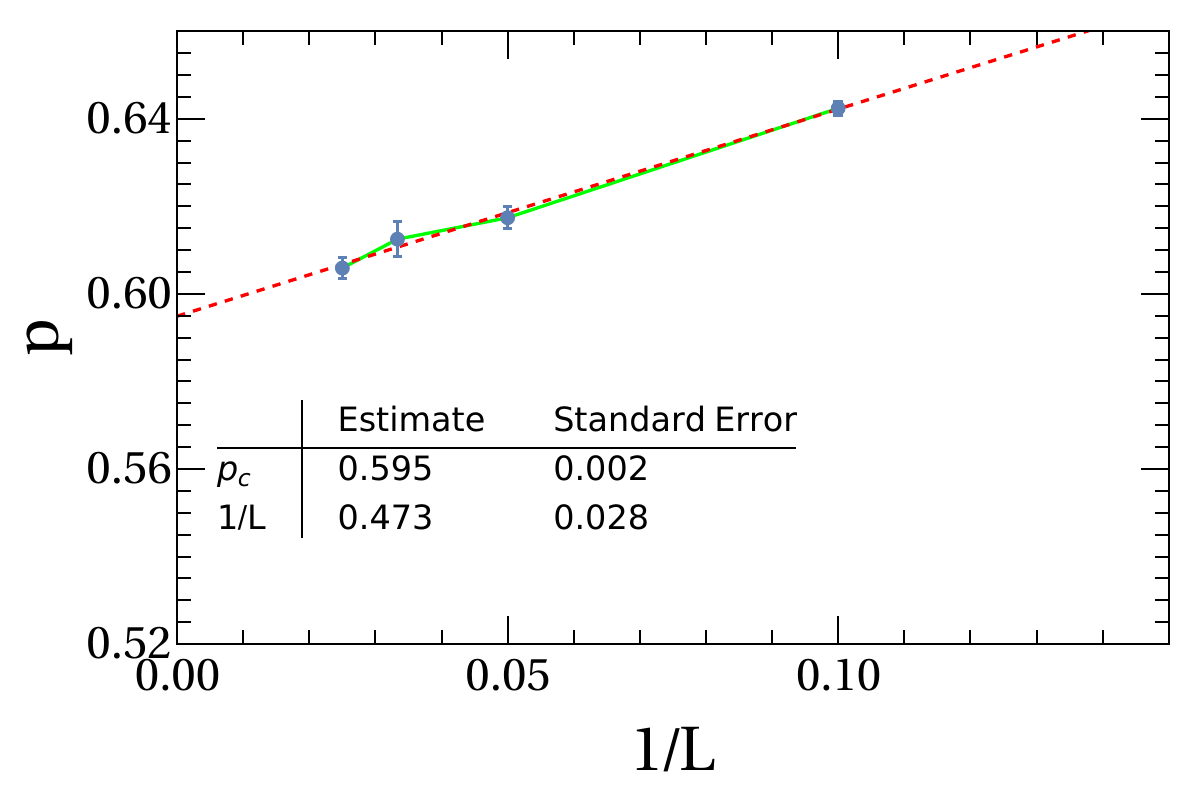} &
    \includegraphics[width=0.30\textwidth]{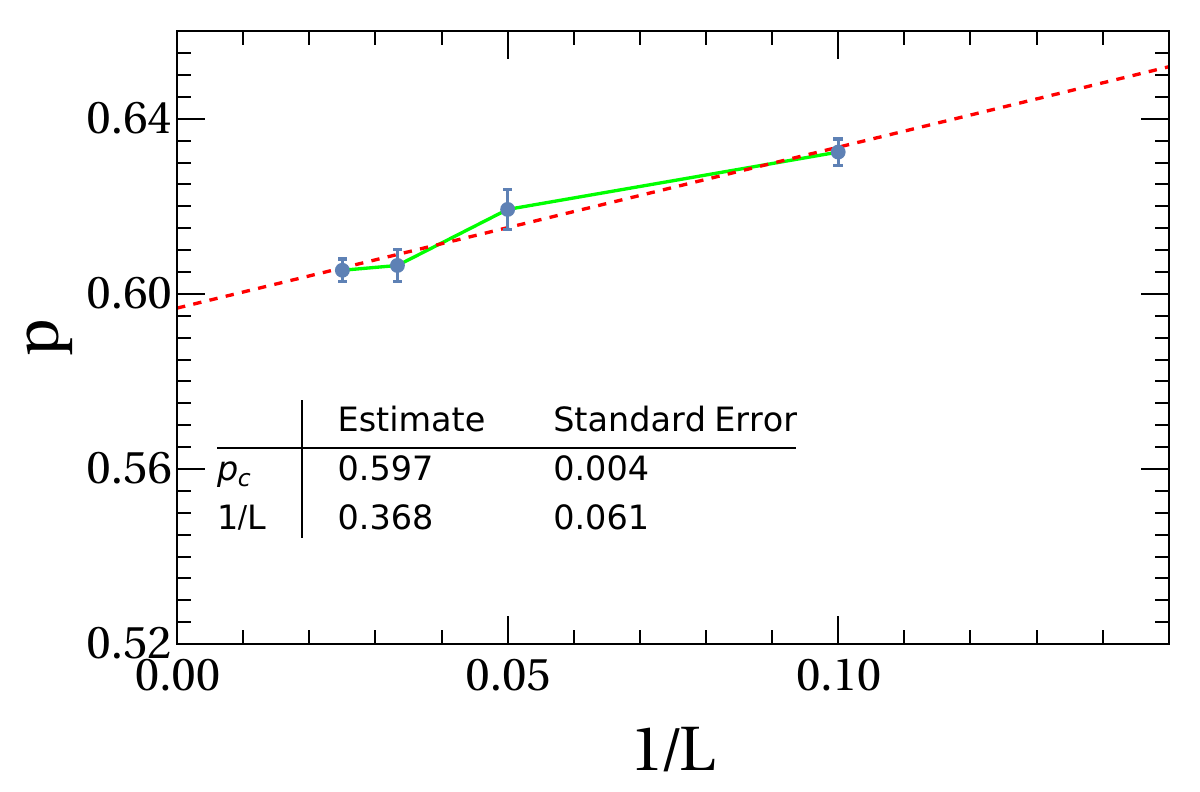}& \includegraphics[width=0.30\textwidth]{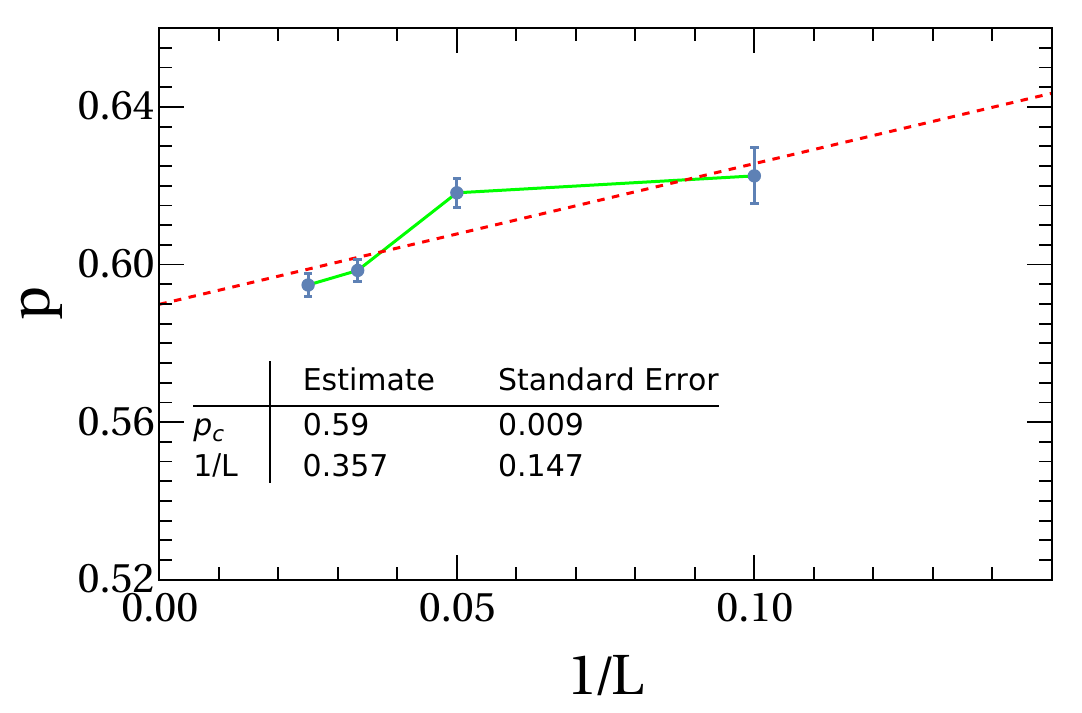} \\
    (d) &  (e) &  (f)  \\
    \includegraphics[width=0.30\textwidth]{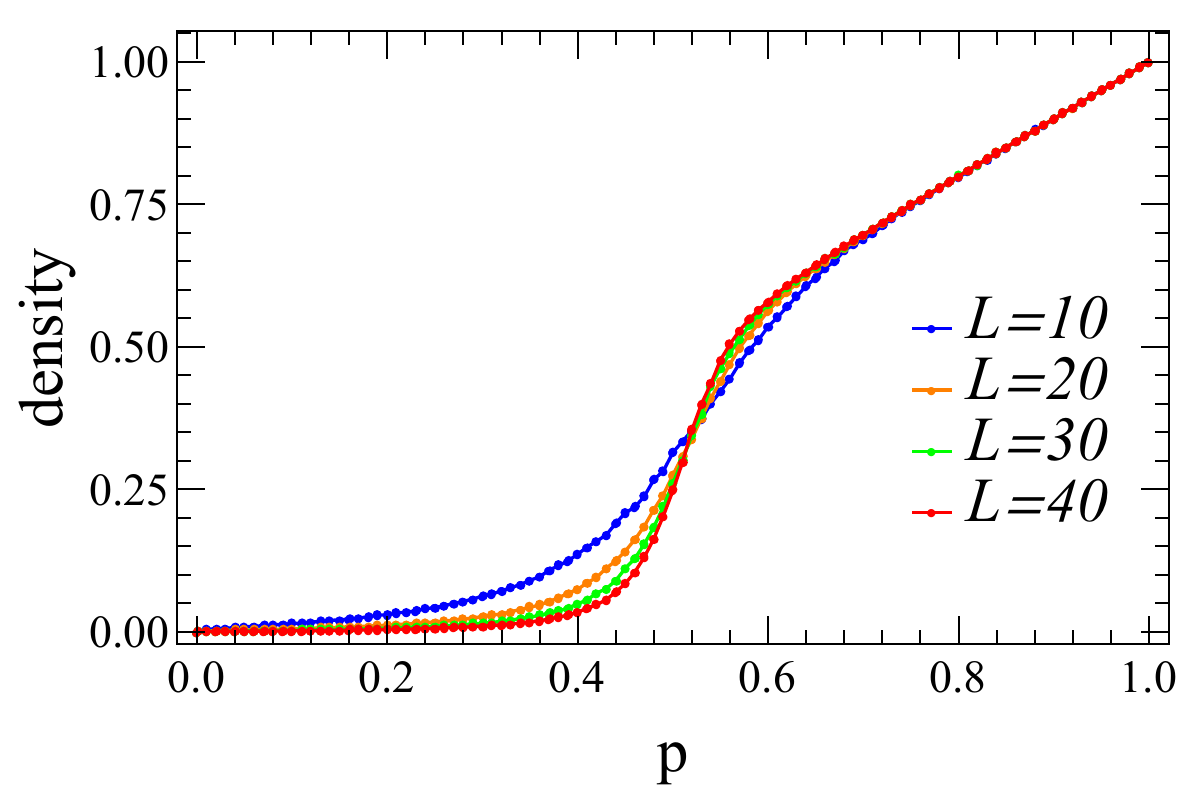}&
    \includegraphics[width=0.30\textwidth]{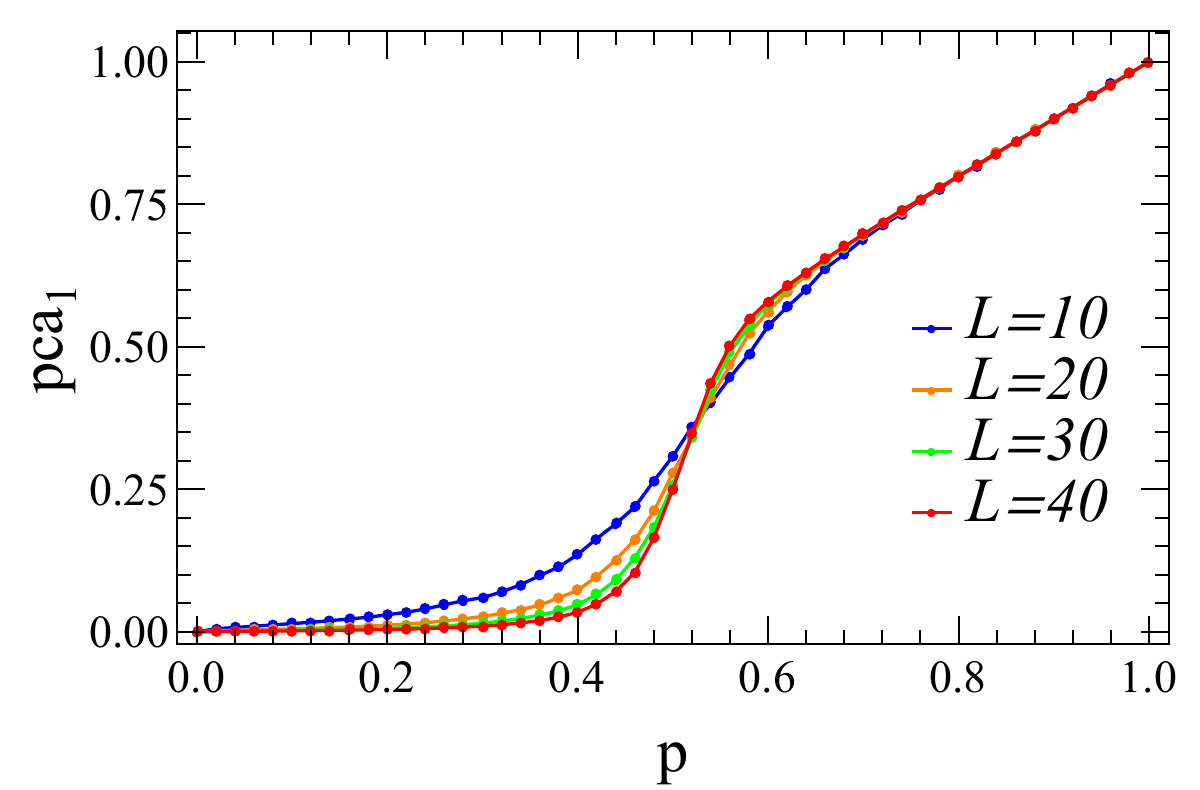}& \includegraphics[width=0.30\textwidth]{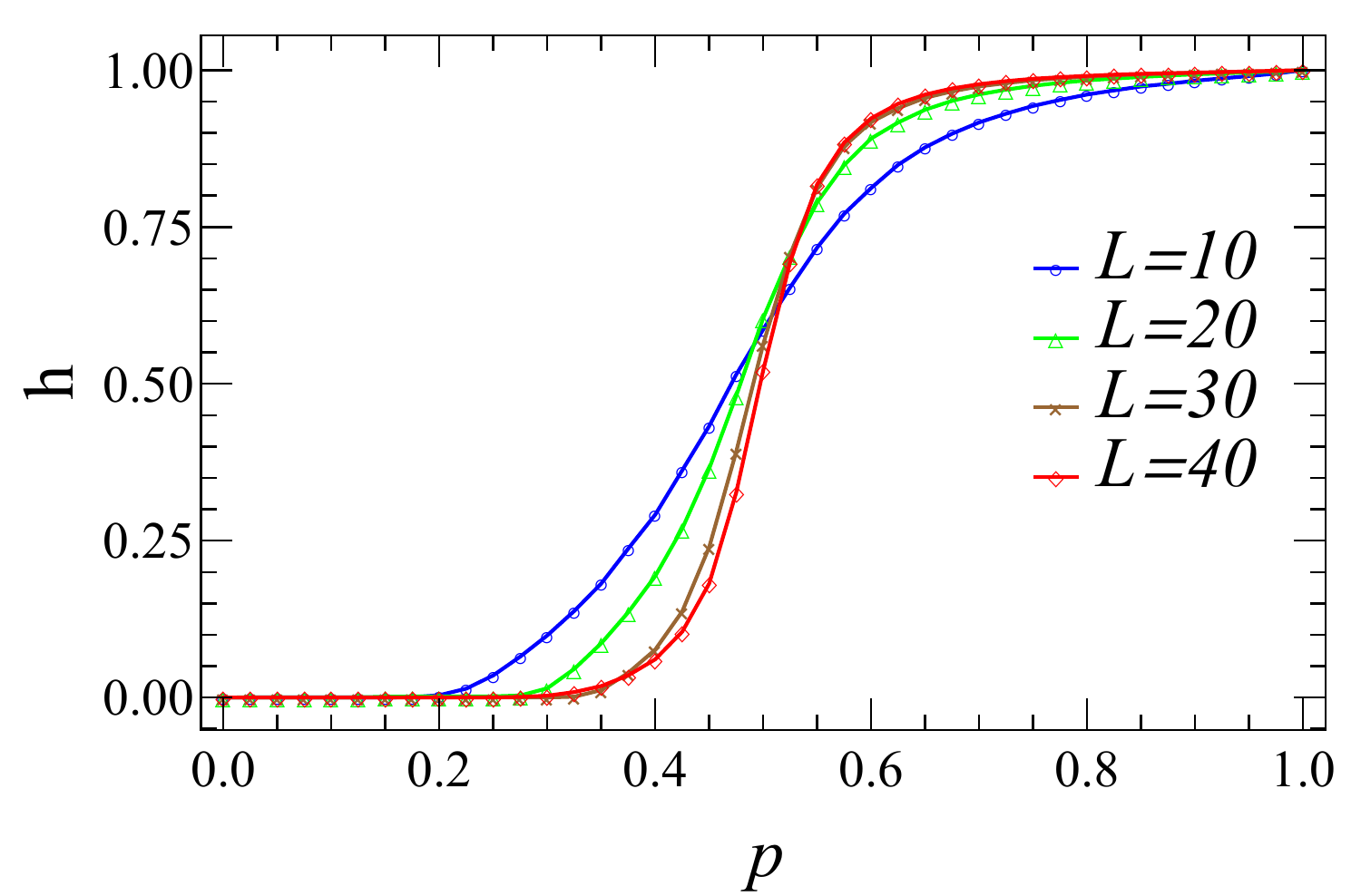} \\
    (g) &  (h) & (i)   \\
    \includegraphics[width=0.30\textwidth]{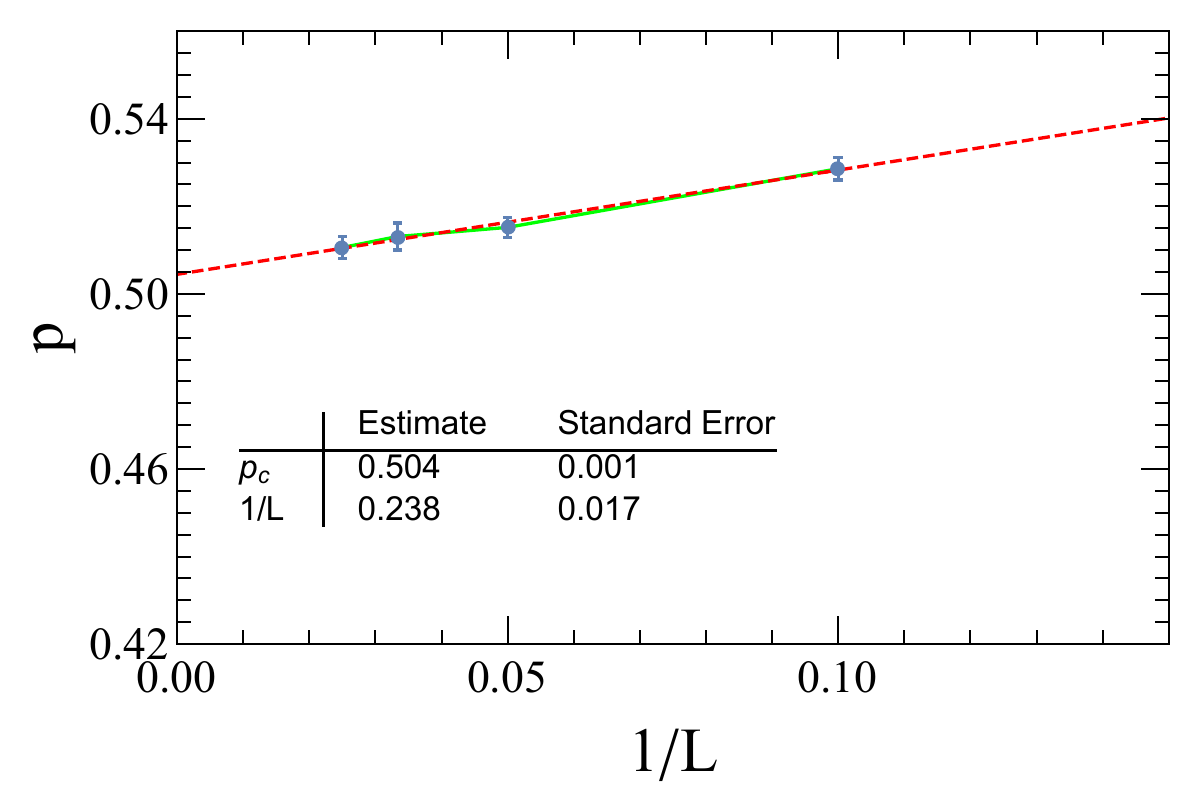} &
    \includegraphics[width=0.30\textwidth]{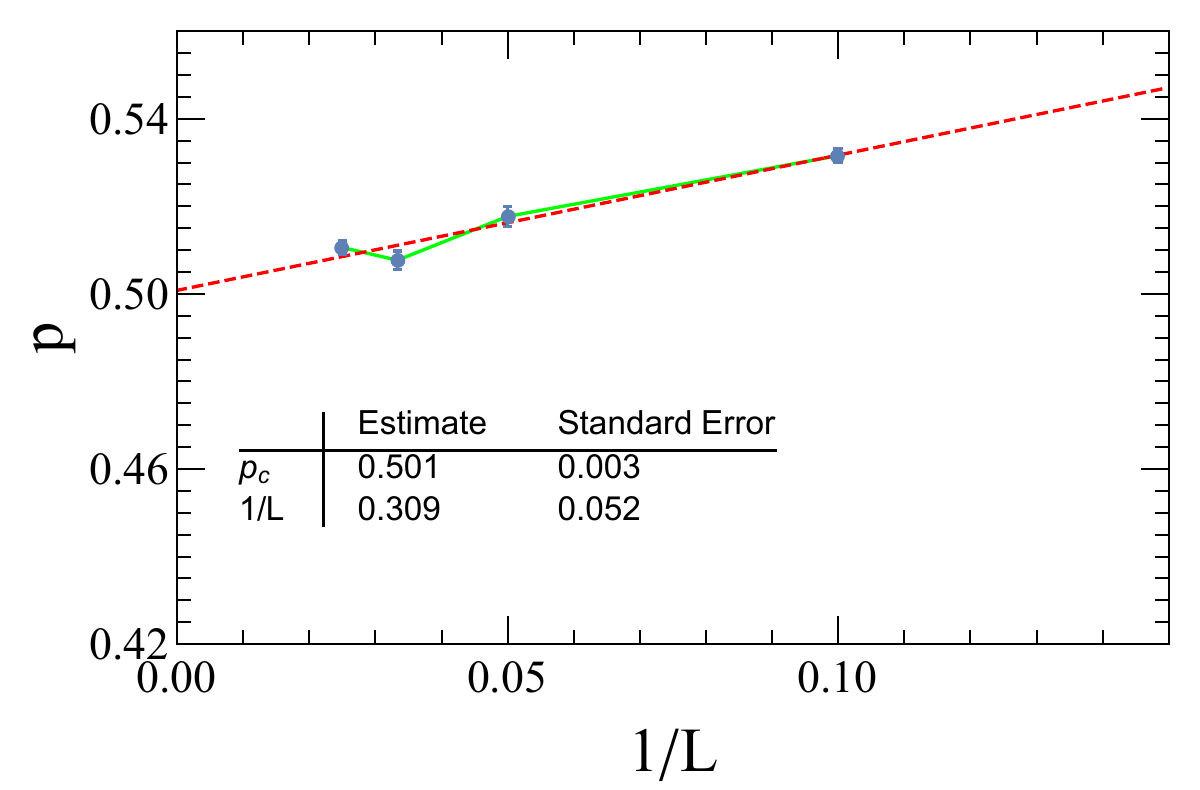}& \includegraphics[width=0.30\textwidth]{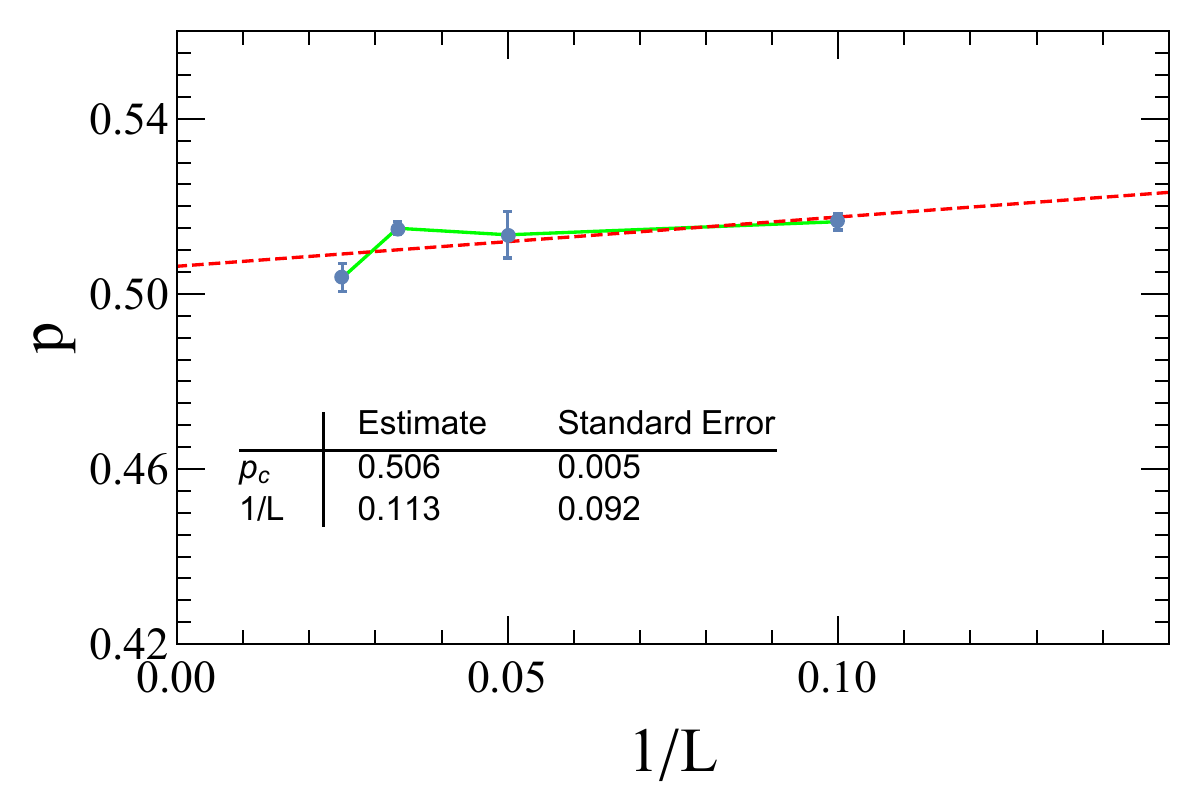} \\
    (j) &  (k) &  (l)
\end{tabular}
\caption{The results of two-dimensional percolation model with the largest cluster selected from the raw configuration. The vertical coordinates of Panels \textbf{a-c} denote the density of active sites, the first principal component of PCA, and the single latent variable of AE, respectively. Their horizontal coordinates are all occupation probability. Panels \textbf{d-f} then correspond to FFSS, FFSS and FSS, respectively. The mean correlation coefficients of $h$, $pca_1$, to density are $0.878\pm(0.003)$ and $0.999\pm(0.001)$, respectively. The vertical coordinates of Panels \textbf{g-i} denote the density of active bonds, the first principal component of PCA, and the single latent variable of AE, respectively. Their horizontal coordinates are all occupation probability. Panels \textbf{j-l} then correspond to FFSS, FFSS and FSS, respectively. The mean correlation coefficients of $h$, $pca_1$, to density are $0.848\pm(0.002)$ and $0.998\pm(0.001)$, respectively.Both the ffss method and the fss method exhibit excellent properties in locating critical points.}
\label{maxtree}
\end{figure*}

In the preceding discussion, we extensively examined the experimental results pertaining to the raw configurations in MC, PCA, and AE. Now, let's delve into the configurations associated with the order parameter in the percolation model, specifically focusing on the largest cluster configuration. As previously mentioned, the occurrence of phase transition in this model predominantly manifests within the largest clusters. 

Faced with the nonlinear curves depicted in Figures \ref{maxtree}(a) and \ref{maxtree}(g), we encounter the challenge of efficiently fitting these data points. To quantitatively derive a reasonable threshold value, we propose a method akin to Finite Size Scaling (FSS), which we tentatively name Fake Finite Size Scaling (FFSS).

The method proceeds as follows: first, we generated the relationship between density of active sites and occupation probability  for site percolation across five different system sizes, $L = 10, 20, 30, 40, 50$, with 1000 samples for each size. Using Mathematica software, we then determined the intersection points $(L_{10-20}, P_{10-20})$ for $L=10$ and $L=20$. Repeating this process, we obtained four intersection points, as listed in Table \ref{FFSS}.

Through FFSS, we extrapolated these intersection points to estimate the system’s behavior as it approaches an infinite size ($\frac{1}{L} \rightarrow 0$). By identifying the intersection of the fitted line with the $p$-axis, we approximated the critical value of site percolation as $p_{c} = 0.595(2)$, as shown in Figure \ref{maxtree}(d), and the percolation threshold for bond percolation as $p_{c} = 0.504(1)$, as shown in Figure \ref{maxtree}(j). Notably, these results align well with theoretical predictions.

Although this method differs from the standard Finite Size Scaling approach, it demonstrates a reasonable degree of validity under the given circumstances.

Simultaneously, Principal Component Analysis (PCA) was applied to analyze the largest clusters in site percolation and bond percolation, yielding the results shown in Figures \ref{maxtree}(b) and \ref{maxtree}(h), respectively. For the PCA results, we observed a functional form similar to that found in Monte Carlo simulation outcomes. To determine the critical points, we employed the Fake Finite Size Scaling (FFSS) method.

The corresponding results are shown in Figures \ref{maxtree}(e) and \ref{maxtree}(k). The extrapolated values in the thermodynamic limit are $p_{c}=0.597(4)$ for site percolation and $p_{c}=0.501(3)$ for bond percolation, which are consistent with theoretical expectations.

In comparison, the results obtained by extracting a single latent variable ($h$) using an Autoencoder (AE) are shown in Figures \ref{maxtree}(c) and \ref{maxtree}(i). The relationship between $h$ and $p$ conforms to a standard Sigmoid function. In this case, we directly used Mathematica to fit the data with a sigmoid function, see equation \ref{sigmoid function}, yielding a critical value of $p_{c}=0.590(9)$ for site percolation, and $p_{c}=0.506(5)$ for bond, which is consistent with the theoretical critical value.
\begin{equation}
    f(x)=\frac{1}{1+e^{-k(x-x_{0})}}
    \label{sigmoid function}
\end{equation}

The finite-size scaling (FSS) results are shown in Figures \ref{maxtree}(f) and \ref{maxtree}(l). Additionally, we calculated the Pearson correlation coefficient between the AE output and Monte Carlo (MC) results, which is 0.878 ± (0.003) for site and 0.848 ± (0.002) for bond. This indicates a strong positive correlation with the density of active sites/bonds. Similarly, we calculated the Pearson correlation coefficient between PCA and MC results, which is 0.999 ± (0.001) and 0.998 ± (0.001). This extremely high positive correlation is evident and can even be directly observed from the figures.

Thus, using the same methodology, the results for site percolation and bond percolation are consistent with theoretical values. This strongly indicates that the methods proposed in this study are valid and effective.

\subsubsection{The shuffled largest cluster}
\begin{figure*}[t]
\begin{tabular}{cccc}
    \includegraphics[width=0.30\textwidth]{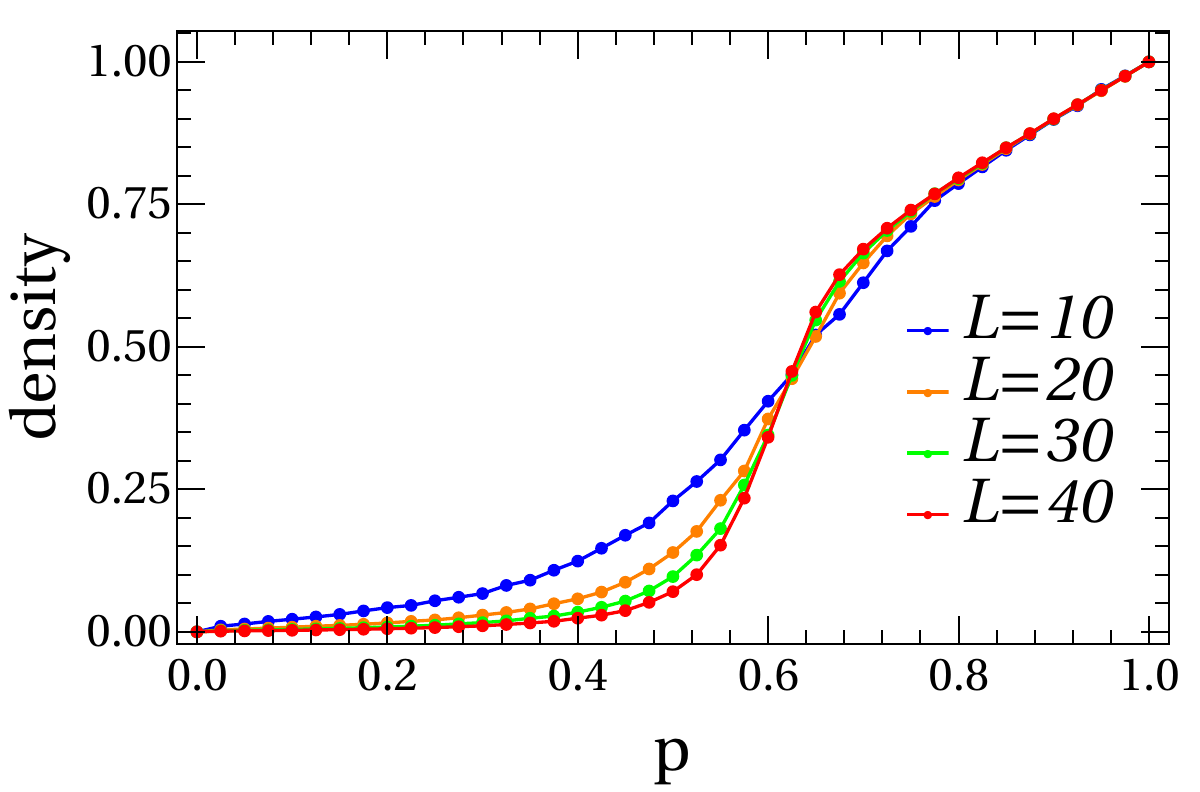}&
    \includegraphics[width=0.30\textwidth]{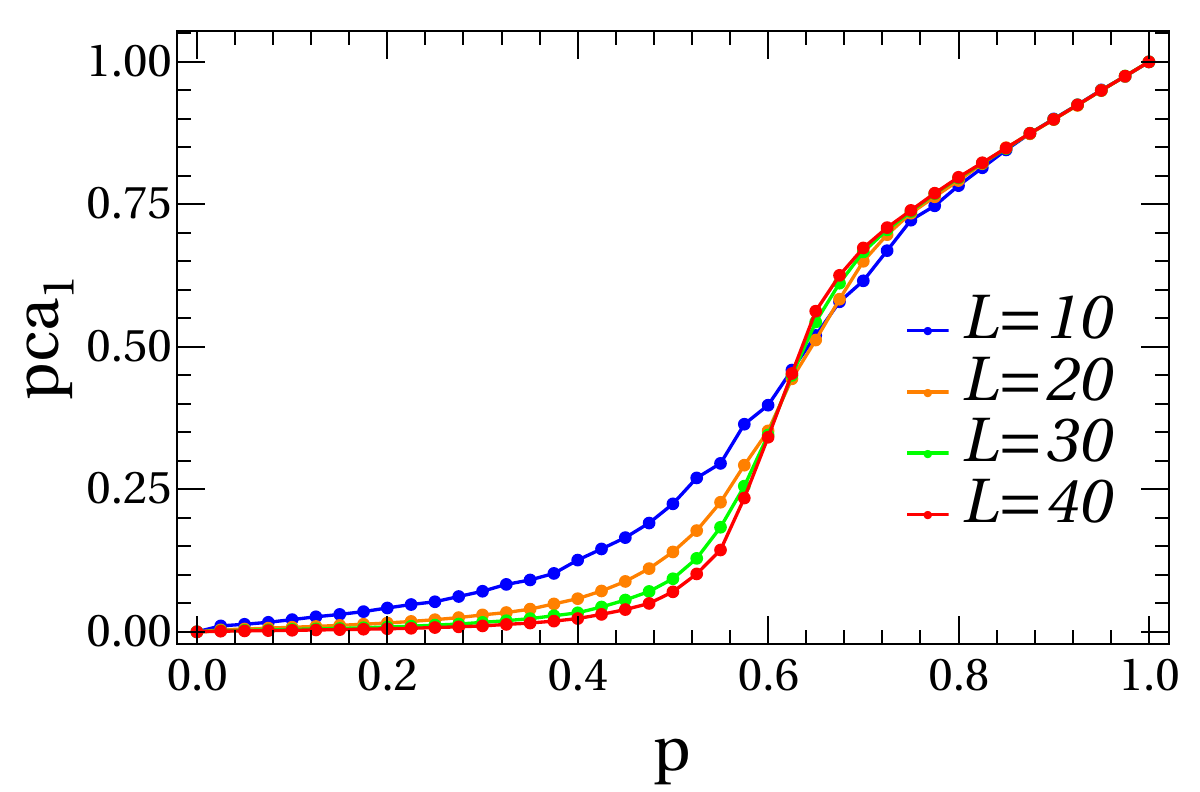}& \includegraphics[width=0.30\textwidth]{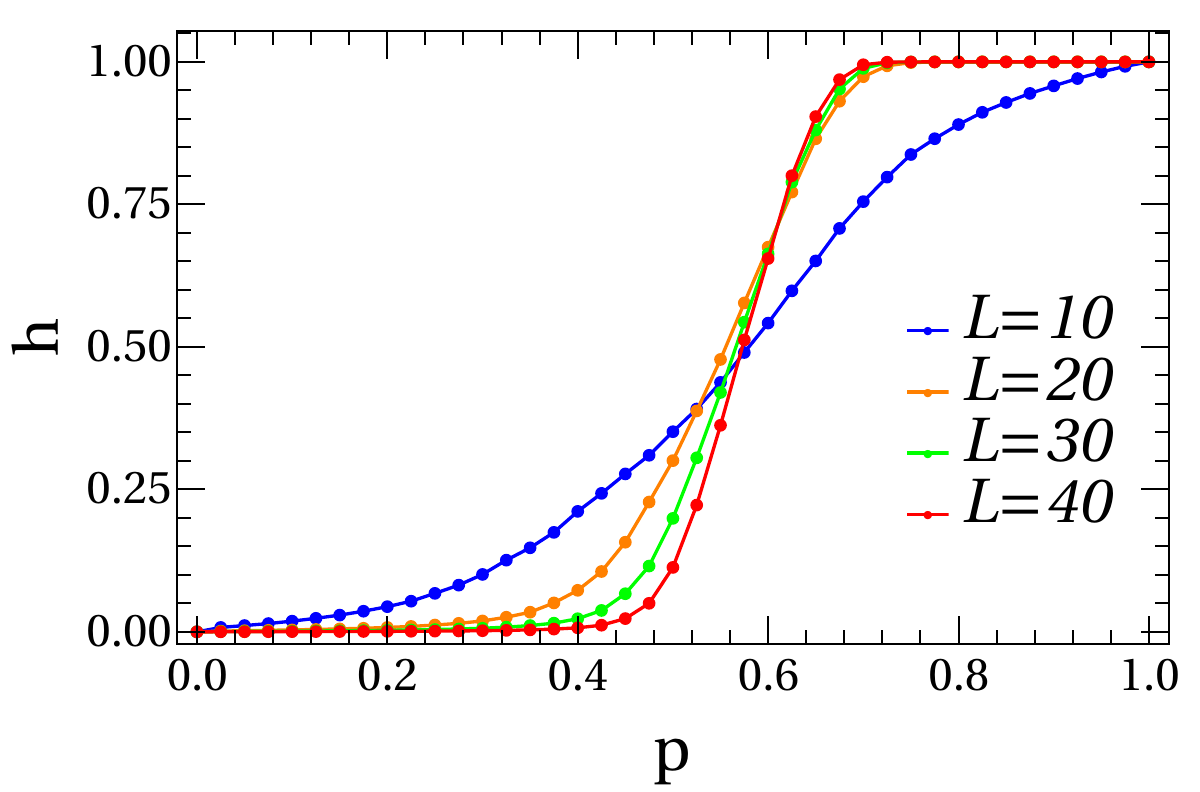} \\
    (a) &  (b) & (c)   \\
    \includegraphics[width=0.30\textwidth]{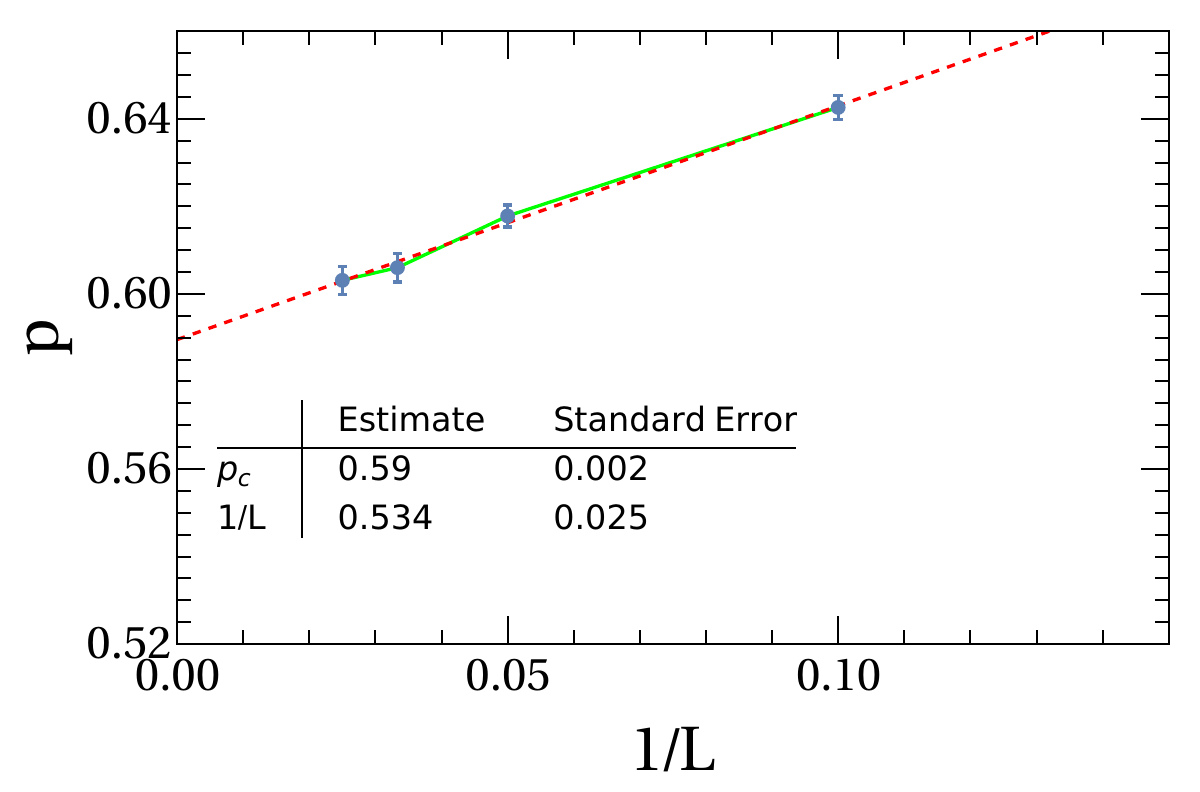} &
    \includegraphics[width=0.30\textwidth]{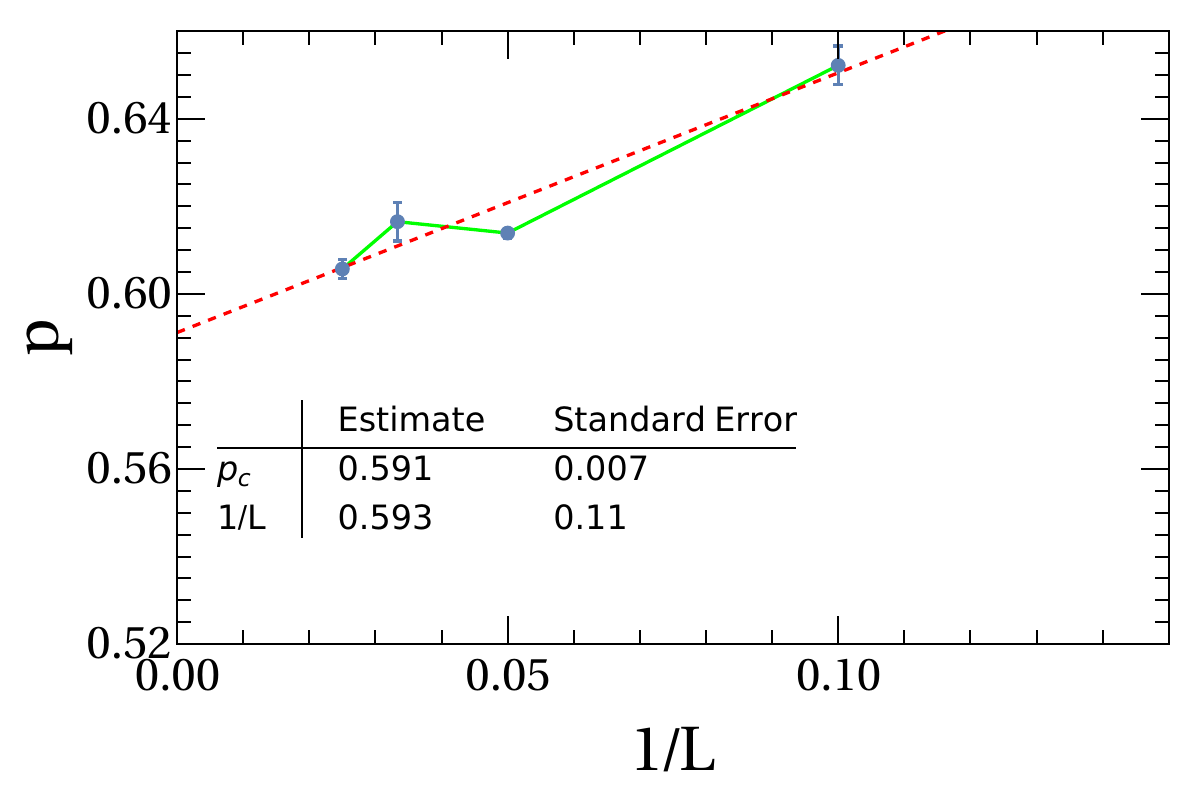}& \includegraphics[width=0.30\textwidth]{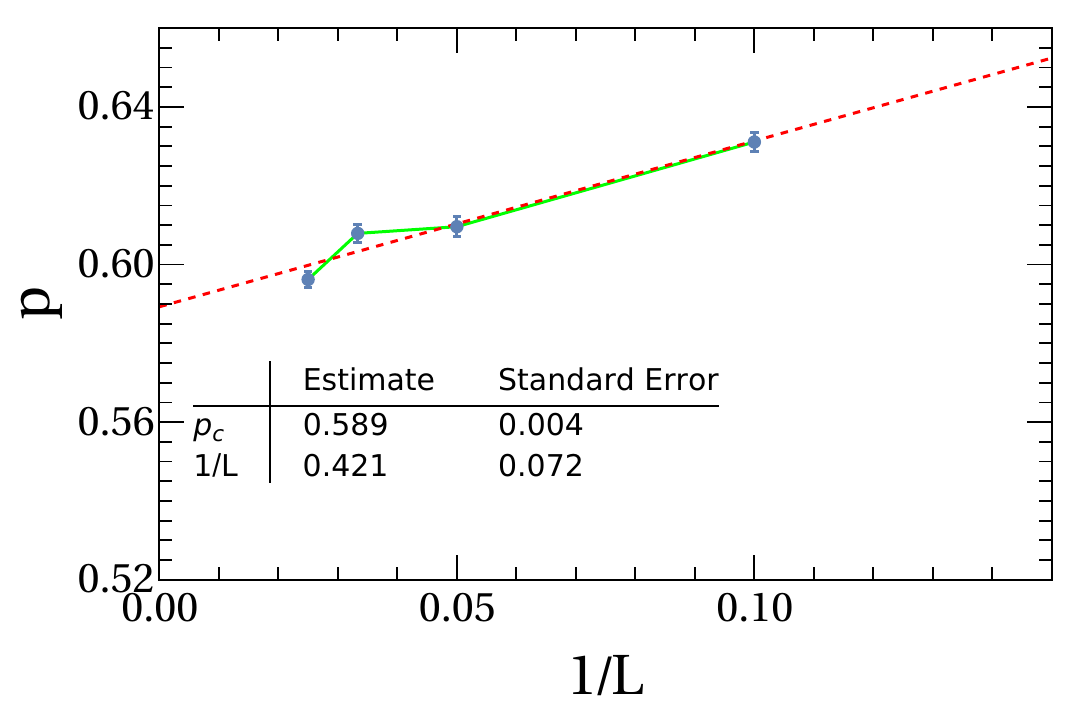} \\
    (d) &  (e) &  (f)
\end{tabular}
\caption{The results of two-dimensional site percolation with the shuffled largest cluster at ratio 0.2. The vertical coordinates of Panels \textbf{a-c} denote the density of active sites, the first principal component of PCA, and the single latent variable of AE, respectively. Their horizontal coordinates are all occupation probability. Panels \textbf{d-f} then correspond to FFSS, FFSS and FSS, respectively. The mean correlation coefficients of $h$, $pca_1$, to density are $0.859\pm(0.004)$ and $0.997\pm(0.001)$, respectively. It exhibits no discernible difference from Figure \ref{maxtree} \textbf{a-f}, whether in terms of locating critical points or the properties of the function's graph.}
\label{maxtree_shuffle}
\end{figure*}

Upon reviewing the outcomes derived from analyzing both the raw configuration and the largest cluster, we have delineated two plausible explanations for the observed results. Firstly, it appears that the largest cluster contains critical information of significance. Secondly, there exists the possibility that unsupervised learning methodologies are solely capable of calculating the density of active sites within the configuration. In order to confirm these hypotheses, we carry out a shuffling experiment on the largest cluster, subject to the performance of the computer, here we only do the above operation on the site percolation model. Should we attain similar results as those obtained with the unshuffled largest cluster (depicted in FIG.~\ref{The largest cluster}) using randomly shuffled largest clusters, it would corroborate that the first principal component of PCA and the single latent variable of the AE indeed encapsulate the density of active sites. Conversely, disparate outcomes would imply that they represent alternative facets of the system.

Utilizing a random shuffling process with a ratio of $0.2$, we perturbed the largest cluster, as illustrated in Figure \ref{percolation_configuration}, during the transition from \textbf{b} to \textbf{c}, while maintaining the constant total number of occupied lattice points, we randomly select twenty percent. of the lattice points (including both active and inactive points) from the entire configuration map for exchange.
and employed the altered configuration to conduct experiments utilizing the three aforementioned methods. The results are illustrated in FIG.~\ref{maxtree_shuffle}. Upon comparison of FIG.~\ref{maxtree_shuffle}(a) with FIG.~\ref{maxtree}(a), it becomes evident from the MC experiment outcomes that the density of active sites remains unaltered. Intriguingly, akin to FIG.~\ref{maxtree}(b), similar outcomes are observed in FIG.~\ref{maxtree_shuffle}(b), with negligible alterations in shape and the critical point persisting around $0.593$. Similarly, FIG.~\ref{maxtree_shuffle}(c) yields outcomes analogous to those of FIG.~\ref{maxtree}(c). This preliminary evidence indicates that the PCA and AE methods exclusively provide insights into the density of active sites of the system phase.

\subsubsection{The largest cluster of shuffled largest cluster}
\begin{figure*}[tbh!]
\begin{tabular}{ccc}   
    \includegraphics[width=0.30\textwidth]{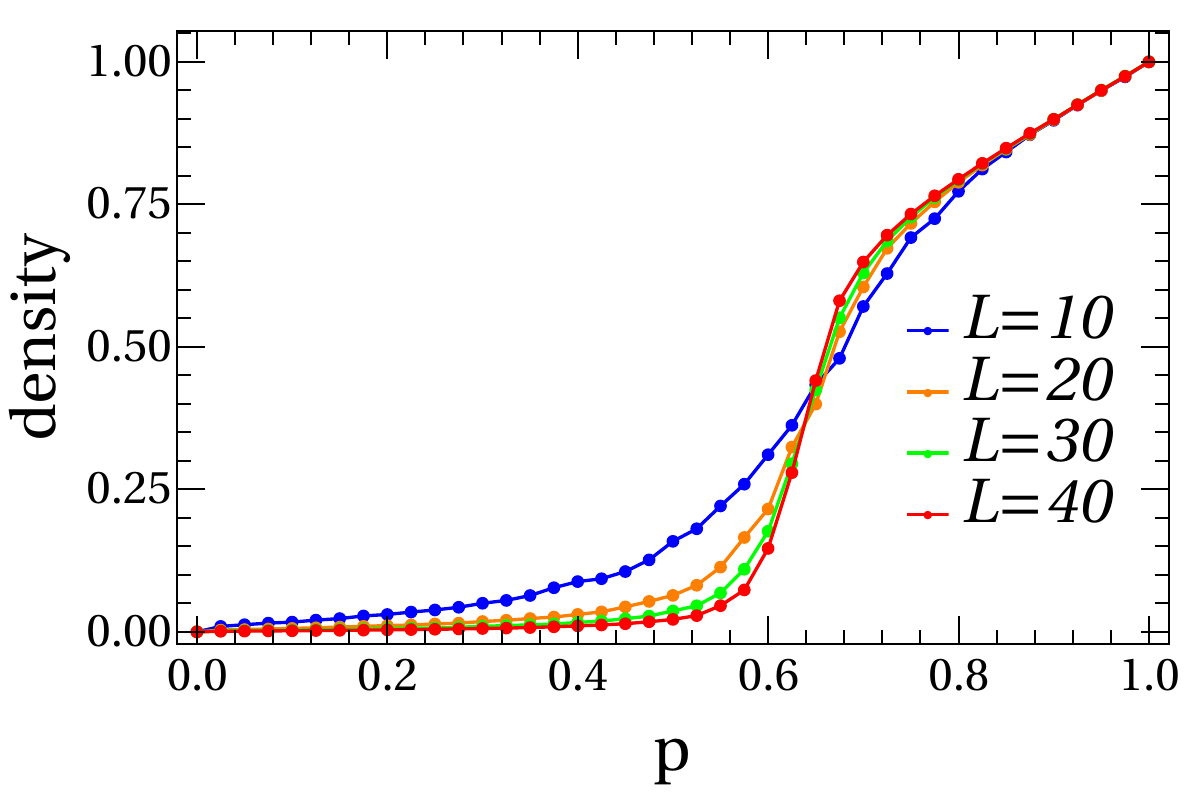} &
    \includegraphics[width=0.30\textwidth]{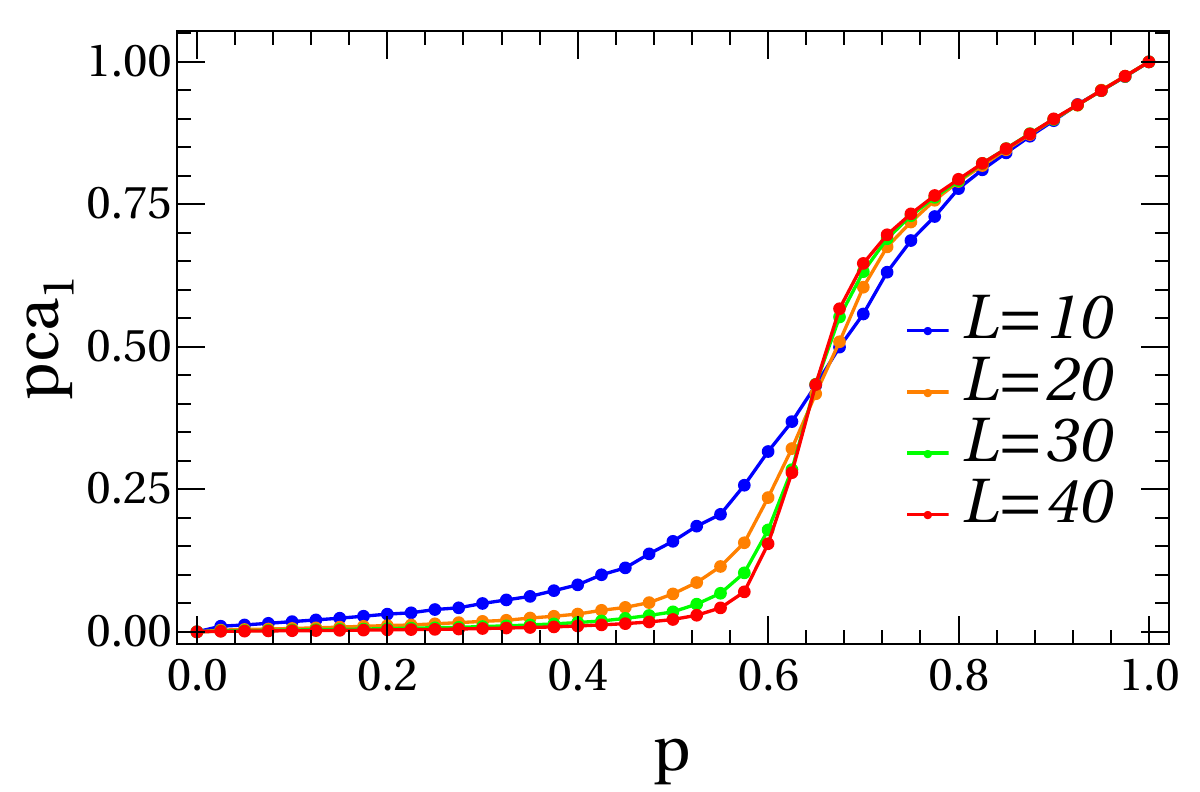}&
    \includegraphics[width=0.30\textwidth]{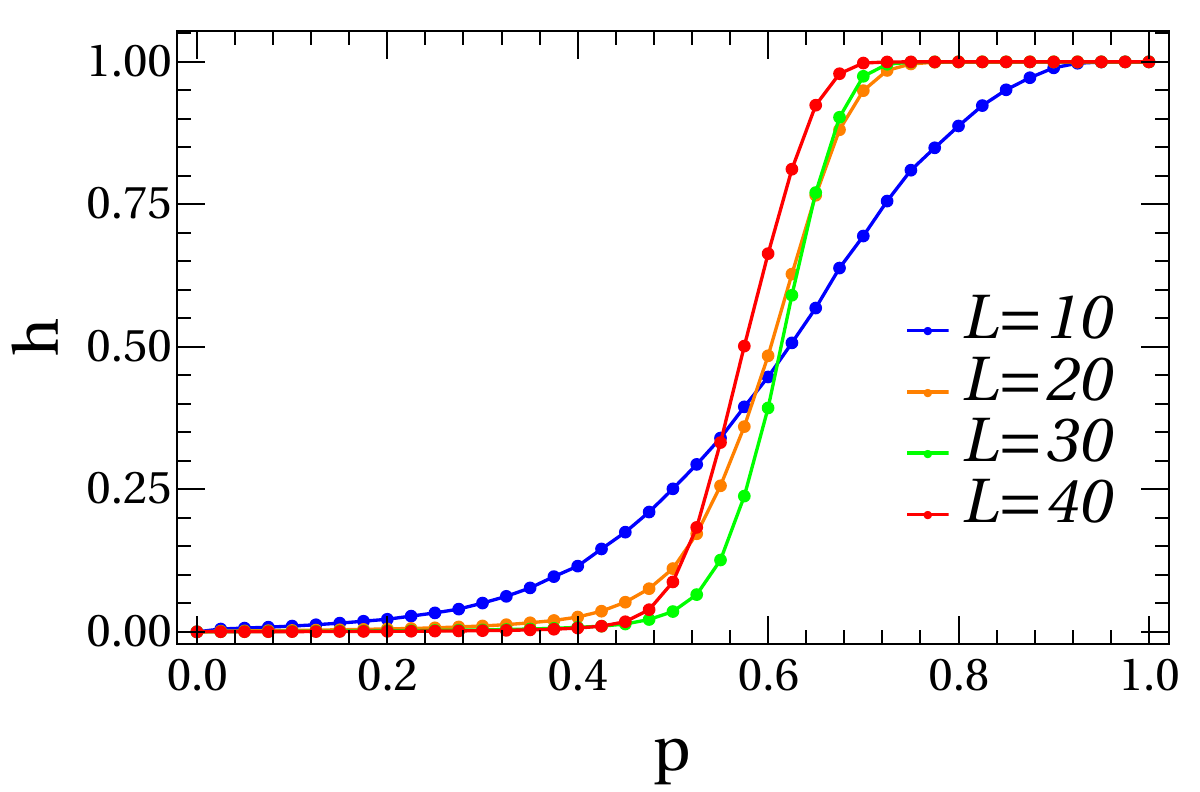}\\
     (a) &  (b) &  (c)
\end{tabular} 
\caption{The experimental results involve selecting the largest cluster from the shuffled largest cluster. panel \textbf{(a)} shows the results of the MC simulation, panel \textbf{(b)} the results of PCA, and panel \textbf{(c)} the results of AE.The mean correlation coefficients of $h$, $pca_1$, to density are $0.849\pm(0.003)$ and $0.988\pm(0.002)$, respectively.A clear observation can be made that the jumping location of the image have shifted to the right.}
\label{maxtree_shuffle_maxtree}
\end{figure*}

Lastly, we extracted the largest cluster from the shuffled configuration to diminish the number of active sites in the configuration while retaining the information pertaining to the largest cluster. This serves as a secondary method to validate our hypothesis. If we procure akin learning outcomes as those derived from the unaltered largest cluster, it indicates that ML can discern certain order parameter information from the largest cluster. Conversely, disparate learning outcomes would imply that the results obtained through ML are inherently linked to density of active sites information.

The results of the MC simulation are depicted in FIG.~\ref{maxtree_shuffle_maxtree}(a). At this juncture, we have significantly altered the original configuration by first extracting the largest cluster and subsequently shuffling it with a certain ratio. Consequently, the spatial correlation of the system has undergone modification. Thus, discussing the phase transition points of the system may no longer be appropriate. However, to encapsulate the overall alteration in this configuration, we opted to employ the term "jumping location". As evidenced by FIG.~\ref{maxtree_shuffle_maxtree}(a), the jumping location appears to shift towards the right. This phenomenon can be attributed to the reduction in density of active sites of the extracted largest cluster following the shuffled configuration. 

FIG.~\ref{maxtree_shuffle_maxtree}(b) showcases the outcome of the first principal component of PCA. Notably, the jumping location of the curve also shifts towards the right, mirroring the trend observed in the MC results. The remarkable concurrence between the MC density of active sites outcomes and the first principal component of PCA lends credence to the notion that the primary information encapsulated by the first principal component of PCA pertains to the density of active sites. This observation underscores that PCA predominantly captures the density of active sites, rather than the specific arrangement of active sites. Once again, it is demonstrated that PCA learns the density of active sites rather than the order parameter.

To further substantiate this perspective, we conducted learning experiments using an AE neural network. FIG.~\ref{maxtree_shuffle_maxtree}(c) illustrates the learning outcome of the AE's single latent variable. Notably, the jumping location depicted in FIG.~\ref{maxtree_shuffle_maxtree}(c) does not exhibit a pronounced shift towards the right. This discrepancy could potentially be attributed to the small shuffle ratio ($r=0.2$).

\begin{figure*}[!htb]
\begin{tabular}{ccc}   
    \includegraphics[width=0.45\textwidth]{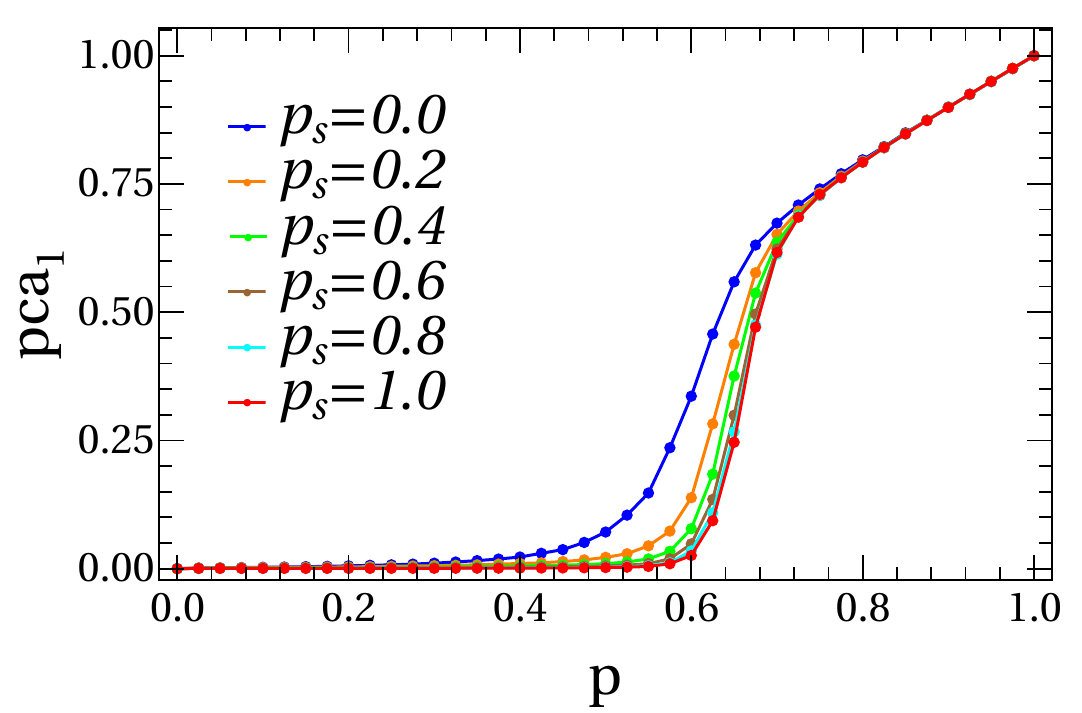} &
    \includegraphics[width=0.45\textwidth]{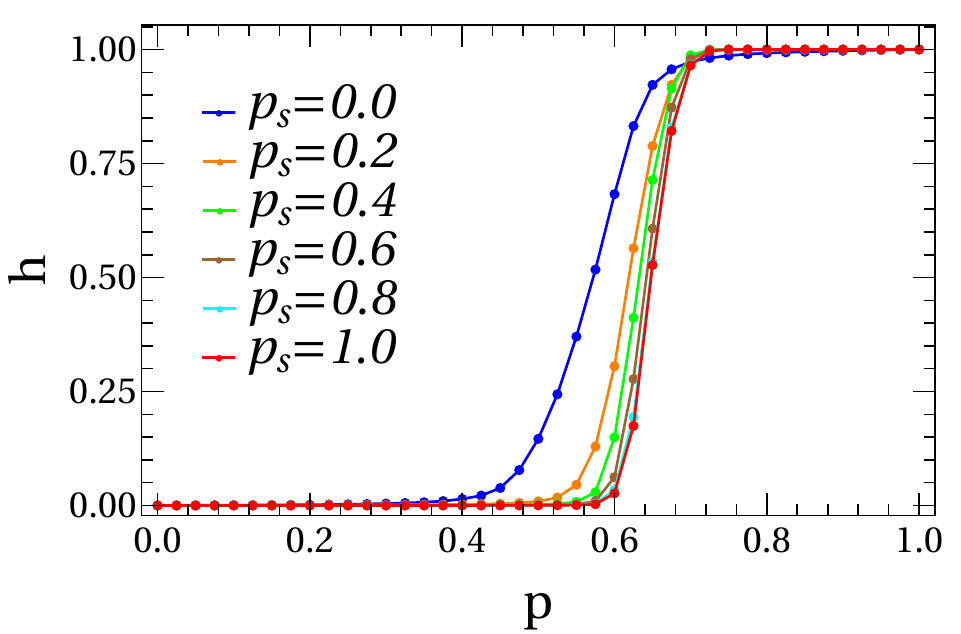}\\
     (a) &  (b)
\end{tabular} 
\caption{The experimental results of selecting the largest cluster from the raw configuration and shuffling it to select the largest cluster conformations are obtained. The system size is fixed at $L=40$, and different shuffling ratios are used for the learning results. Panel \textbf{a} shows the results of PCA, and panel \textbf{b} shows the results of AE. With an increase in the probability of shuffling, it is evident that the jumping location of the images are shifting progressively to the right.}
\label{maxtree_shuffle_maxtree_percent}
\end{figure*}

In order to delve into the impact of the shuffle ratio on the jumping location in greater detail, we maintained the system size at $L=40$ and shuffled the largest cluster at varying ratios, namely $0$, $0.2$, $0.4$, $0.6$, $0.8$, and $1$. These modified configurations were then fed into PCA and autoencoder for learning, respectively, as depicted in FIG.~\ref{maxtree_shuffle_maxtree_percent}. FIG.~\ref{maxtree_shuffle_maxtree_percent}(a) displays the PCA results, wherein a discernible rightward shift in the jumping location of the curve is evident with increasing shuffling ratio when the system size remains constant. Similarly, FIG.~\ref{maxtree_shuffle_maxtree_percent}(b) presents the results obtained from the AE. Analogous to PCA, an unmistakable rightward shift in the jumping location of the function is observed as the shuffling ratio escalates. This further underscores that both methods primarily capture the density of active sites of the configuration rather than the order parameter. For more detailed AE results, refer to TABLE.~\ref{ae_shuffle_ratios}. The jumping location in TABLE.~\ref{ae_shuffle_ratios} is determined through sigmoid function fitting. It is evident that as the shuffling ratio increases, the jumping location also increases, underscoring the robustness of the AE's single latent variable results.

\begin{table*}[tbh!]
	\centering
	\resizebox{\textwidth}{7mm}{
	\begin{tabular}{|c|c|c|c|c|c|c|c|c|c|c|c|}
        \hline
  $r$(shuffle ratio)   & 0  &0.1 &0.2 &0.3 &0.4 &0.5&0.6&0.7&0.8&0.9&1.0 \\
        \hline
   $jumping~location$ & 0.5695(6) &0.6020(4)  &0.6192(2)  &0.6274(2)  &  0.6323(2)&0.6384(2) &0.6423(2)&0.6454(1)&0.6482(2)&0.6491(2)&0.6491(2) \\
        \hline
   \end{tabular}}
\caption{AE results of two-dimensional site percolation with different shuffle ratios, where $L=40$. The second row in the table represents the single potential variable results of the largest cluster after shuffling the largest cluster with different shuffle ratios. Corresponding to Figure\ref{maxtree_shuffle_maxtree_percent}\textbf{b}, the sigmoid function is employed.}
\label{ae_shuffle_ratios}
\end{table*}

\subsection{Discussions}\label{Discussions}
\begin{table*}[tbh!]
	\centering
	\begin{tabular}{|c|c|c|c|}
        \hline
       \diagbox[width=10em]{method}{$p_{c}$}{configuration}     & raw  &largest cluster &shuffled largest cluster     \\
        \hline
   MC($density$)  & None &0.595(2)        &0.590(2)     \\
        \hline
   PCA($pca_{1}$) & None &0.597(4)        &0.591(7)     \\
        \hline
   AE($latent$)   & None &0.590(9)        &0.589(4)     \\
        \hline
   \end{tabular}
\caption{This table presents the critical values of the two-dimensional site percolation model ($L = 40$) under three different methods, MC, PCA, and AE, where the shuffle ratio of the largest cluster is $r = 0.2$.}
\label{discussion}
\end{table*}

The four parts above respectively explore four different configurations of site percolation using MC simulations and unsupervised learning techniques: the raw configuration, the largest cluster, the shuffled largest cluster, and the largest cluster after shuffling. These findings are summarized in TABLE.~\ref{discussion}. The raw configuration of site percolation is randomly generated, and the density of active sites does not reveal critical information, thus no critical point is identified in this column. In the largest cluster column of TABLE.~\ref{discussion}, we employ three methods to ascertain the critical value. The results indicate that the largest cluster can effectively represent the critical information of the site percolation model. This observation motivates further investigation into the relationship between density of active sites, PCA's first principal component, AE's individual latent variables, and order parameters.

In the shuffled largest cluster column of TABLE.~\ref{discussion}, all three methods still capture the critical point of the model effectively. This suggests that PCA's first principal component and AE's single latent variable precisely extract the number of active sites in the system, which appears to have little association with the specific spatial arrangement of the active sites. Lastly, a similar study is conducted on the largest cluster after shuffling. Due to the altered spatial correlation of the system, discussion shifts from the system's phase transition point to the jumping location, illustrating the influence of shuffle ratio.

\section{Conclusion}\label{Conclusion}

Previous attempts to locate the critical point of percolation models using unsupervised learning on raw configurations have been unsuccessful. In this paper, we propose using the largest cluster to identify the critical point and achieve promising results with Monte Carlo (MC) simulations, Principal Component Analysis (PCA), and Autoencoders (AE). Additionally, we introduce a novel FFSS method that significantly aids in extracting the critical point from unconventional functional images.

In the process of capturing the critical point using largest cluster, we observed that the PCA and AE learning results of the raw and largest clustering configurations were very similar to the density of active sites MC results. Through a thorough examination of various model configurations, we hypothesized that the first principal component of PCA and the single latent variable of AE may inherently reflect density of active sites. To test this hypothesis, we performed a random shuffle of the largest cluster and found that the experimental results did not change significantly before and after the shuffle. This confirms that the first principal component of PCA and AE’s single latent variable effectively represents the density of active sites in the percolation model.

However, we noted that for largest clusters of the same size, different shuffling probabilities affected the size of the remaining largest clusters. By analyzing the remaining largest clusters after varying degrees of shuffling, we observed a shift in the system's inflection point. This indicates that random shuffling alters the correlation length of the system—the greater the shuffling ratio, the smaller the correlation length. This further supports the notion that the first principal component of PCA and the single latent variable of AE have a physical interpretation related to density of active sites.

\section{Acknowledgements}

This work was supported in part by Key Laboratory of Quark and Lepton Physics (MOE), Central China Normal University (Grant No.QLPL2022P01), Research Fund of Baoshan University(BYPY202216, BYKY202305), Wen Bangchun Academician Workstation(202205AF150032), the Fundamental Research Funds for the Central Universities, China (Grant No. CCNU19QN029), the National Natural Science Foundation of China (Grant No. 61873104), the 111 Project, with Grant No. BP0820038.

During the preparation of this work the author used GPT-4 in order to improve language and readability. After using this tool, the author reviewed and edited the content as needed and take full responsibility for the content of the publication.

The detailed algorithms of how to generate data and use machine learning are shown in the GitHub link
{https://github.com/freeupcoming/site-percolation}.

\bibliographystyle{apsrev4-2}
\bibliography{sitepercolation}

\end{document}